\documentstyle[aas2pp4,tighten]{article}

\begin{document}


\title{TILTING OF A DISK OF GRAVITATING RINGS} 
 
\author{R.V.E. Lovelace} 

\affil{Department of Astronomy, 
Cornell University, Ithaca, NY  14853-6801; rvl1@cornell.edu}

\begin{abstract}

 The present work represents an attempt
to understand the `rules of behavior' of
observed warps in the HI disks of spiral
galaxies found by Briggs (1990).
 In contrast with most earlier
theoretical work, the present
study investigates different
initial value problems of
a warped disk in an oblate (or prolate)
halo potential, and it 
represents the disk warp
in terms of $N$ independently
tilted, self-gravitating, concentric
rings.  
 This representation gives
new insight into the disk warping.
 A new constant of the motion of
$N$ tilted rings is identified (in addition
to the energy).
 The phenomenon
of phase-locking of the lines-of-nodes
of nearby rings due to self-gravity
is demonstrated.
 We consider the influence of dynamical friction
due to ring motion through the halo
matter as well as friction between
gaseous rings with different vertical motions
due to turbulent viscosity.
 We first consider the dynamics of
one, two, and three tilted rings
of different radii in a
halo potential.  
 We go on to develop dynamical equations
for $N-$rings which are most simply 
expressed in terms of the complex
tilt angles $\Theta_j = 
\theta_j {\rm exp}(-i\varphi_j)$, where $\theta_j$
is the actual tilt angle and $\varphi_j$ the 
line-of-nodes angle for the $j^{th}$ ring ($j=1..N$).
 We numerically solve the 
equations for
$\Theta_j$ for four different types of initial 
conditions:  
 (1) warp excitation by a 
passing satellite, 
(2) excitation by
a sinking compact minor satellite, 
(3) warp evolution due to
a tilted halo potential, and 
(4) warp evolution
resulting from an initially tilted disk
plane.

\end{abstract}


\section {Introduction}

The large-scale integral-sign 
warps of the outer regions of spiral
galaxies has been a long-standing
theoretical puzzle.  
 The strongest warps are observed in the
HI disks when these extend well beyond the
optical disks (Briggs 1990).  
 The warps are antisymmetric and 
consequently can be well-fitted 
by the kinematic tilted-ring 
model of Rogstad, Lockhart,
and Wright (1974).  
 For a set of 12 galaxies with high-quality
data on the extended HI disks, Briggs (1990)
found empirically a set of `rules of behavior'
of warps: 
1. The warps develop with increasing
radius $r$ for $r$ between 
$R_{25}$ and $R_{Hol}$
(the Holmberg radius, $=R_{26.5}$), 
where the subscript
denotes the surface brightness in $B$ in
mag/arcsec$^2$;
2. The line-of-nodes of the warp is straight
for $r < R_{Hol}$;  and
3. For $r> R_{Hol}$, the line-of-nodes forms
an open leading spiral.
 An illustrative case is shown
in Figure 1 which gives
a surface plot of the warp 
of the disk of the spiral galaxy
M 83 obtained from the data of Briggs (1990).
The rotation of M 83 is clockwise so that the
warp is a {\it leading} spiral wave. 
 Figure 2 also shows the data of Briggs (1990)
for M 83, in panel ({\bf a})
the radial ($r$) dependence of the 
tilt angle $\theta$, in ({\bf b})
the $r$-dependence of the angle 
of the line-of-nodes $\varphi$,
and in ({\bf c}) the
dependence of $\theta$ on 
$\varphi$, which we refer
to as a Briggs plot.
 A different behavior is observed
for the galaxy NGC 3718
which shows an approximately straight
line-of-nodes out to  
$r\approx 1.75 R_{Hol}$ and a large
tilt angle ($\sim 90^o$) at this
radius (Briggs 1990). 

\placefigure{fig1}

\placefigure{fig2}

 A partial listing of the theoretical
works on warps of disk galaxies include 
Kahn and Woltjer (1959), 
Hunter and Toomre (1969; hereafter HT),
Toomre (1983), 
Sparke and Casertano (1988), 
Binney 1992, Hunter (1994), 
Merritt and Sellwood (1994), 
and Nelson and
Tremaine (1995).
 Since the pioneering analysis by
Hunter and Toomre (1969) and
the earlier related work by
Lynden-Bell (1965), the theoretical
studies have focused mainly on determining
the eigenmodes and eigenfrequencies
of warped self-gravitating disks.
 The studies used the HT
dynamical equation for the vertical
displacement of the disk 
surface $h(r,\phi,t)$, 
where $\phi$ is the
azimuthal angle.

 In contrast with these earlier works,
the present study investigates different
initial value problems of a warped
disk in an oblate (or prolate) halo
potential, and it develops
a different representation for the disk
warping.  
 Instead of describing the
disk in terms of $h(r,\phi,t)$, we 
develop
a representation in terms of $N$
independently tilted, 
self-gravitating, concentric rings
which is suggested by the kinematic
ring model of Rogstad et al. (1974).
 May and James (1984) mention
a model of this kind, but 
they give no details.
 We show that the $N-$ring
representation is equivalent to
the HT equation for $h(r,\phi,t)$
for $N \rightarrow \infty$.
 However, new insight is given by the
$N-$ring model:  
 A new constant of the motion for 
a warped disk is found. 
 The notion of the phase-locking
of the line-of-nodes of nearby
rings due to self-gravity is
demonstrated.  
We propose that this explains Briggs's rule No. 2.
 Furthermore, treatment of initial
value problems for the age of
a galaxy can be done with
sufficient accuracy with relatively
small values of $N$ ($ {\buildrel < \over \sim}10^2$).

 In \S 2 we first give basic
parameters of the disk and
the halo.
 In \S 2.1 
we treat the case of 
one tilted ring and in \S2.2
comment on the interpretation of
observations of a ring. 
 Subsection 2.3 considers the 
case of two tilted rings.
 Subsection 2.4 treats the case 
of two tilted, counter-rotating
rings.
 Subsection 2.5 treats the case of three 
corotating rings.
Section 3 treats the general case of 
$N$ tilted corotating rings.
 In this Section we consider different
possible excitations 
or origins of warps - that due to
a passing satellite (\S 3.1), a sinking
satellite (\S 3.2), a tilted halo potential
(\S 3.3), and an initially tilted outer
disk plane (\S 3.4). 
Section 4 discusses
the continuum limit where $N\rightarrow \infty$.
Section 5 gives conclusions of
this work.

\section {Tilting of Disk Galaxies}

The equilibrium, unwarped galaxy is assumed to be 
axisymmetric and to consist of a thin disk of
stars and gas and a slightly oblate or prolate halo
of dark matter.  We use an inertial cylindrical $(r,\phi,z)$
and Cartesian $(x,y,z)$ coordinate systems with the
disk and halo equatorial planes in the $z=0$ plane.  The
total gravitational potential is written as
$$\Phi(r,z)=\Phi_{id} +\Phi_{od}+\Phi_h~,
\eqno(1)$$
where $\Phi_{id}$ is the potential due to the inner
part of the disk (as discussed below), $\Phi_{od}$ is
that for the outer disk, and $\Phi_h$ is that for the
halo.  The particle orbits in the equilibrium disk
are approximately circular with angular
rotation rate $\Omega(r)$, where
$$\Omega^2(r)={1\over r}
{{\partial \Phi}\over{\partial r}}\bigg|_{z=0} =
\Omega_{id}^2+\Omega_{od}^2 +\Omega_h^2~.
\eqno(2) $$
 The vertical epicyclic
frequency (squared),
$$\kappa_{z}^2(r)\equiv {{\partial^2\Phi}
\over{\partial z^2}}\bigg|_{z=0}
 =\kappa^2_{idz}+\kappa_{odz}^2+\kappa_{hz}^2~,
\eqno(3)$$
gives a measure of the restoring force in the 
$z-$direc-tion.

The surface mass density of the inner (optical) disk 
is assumed to be $\Sigma =\Sigma_0 {\rm exp}
(-r/r_d)$ with $\Sigma_0$ and $r_d$  
constants and $M_d=2\pi r_d^2\Sigma_0$ the
total disk mass.  For $r^2\gg r_d^2$,
$$\Omega_{id}^2 = {{GM_d}\over{r^3}}\left(1+
{9\over2}{{r_d^2}\over{r^2}}\right)~,$$
 $$\kappa_{idz}^2={{GM_d}\over{r^3}}\left(1+
{27\over2}{r_d^2\over r^2}\right)~,
\eqno(4)$$
(Binney and Tremaine 1987; hereafter BT, p. 409).  
 Note that $\kappa_{idz}^2-\Omega_{id}^2
=9GM_dr_d^2/r^5$. 

The outer, neutral hydrogen disk
can be described approximately by a 
Fermi function for the surface
density
$$\Sigma_H={\Sigma_{H0}\over 1+{\rm exp}[k_1(r/R_H-k_2)]}
\eqno(5)
$$
where $R_H$ is the radius inside of which half
the neutral hydrogen mass is located, $k_1\approx 3.70$,
and $k_2\approx 1.21$.
This dependence provides a good fit to
the data of Broeils and Rhee (1966; their Figure 6a).
The total neutral hydrogen mass
is $M_H \approx 5.35 R_H^2\Sigma_{H0}$.

 The halo potential is taken to
be $$\Phi_h = {1\over2}v_o^2 ~
{\ell} n\left[r_o^2+r^2+(z/q)^2\right]~,
\eqno(6a)$$
where
$v_o=$ const. is the circular 
velocity at large distances, 
$r_o=$ const. is the core radius 
of the halo, and $q=const.$ is the axial
ratio of the equipotential surfaces 
with $\varepsilon \equiv  1-q$
their ellipticity.  An oblate (prolate) 
halo corresponds to  $q<1$ ($q>1$).
We have 
$$\Omega_h^2 = {{v_o^2}\over{r_o^2+r^2}}~,~~~~ 
\kappa_{hz}^2={{(v_o/q)^2}\over
{r_o^2+r^2}}~.
\eqno(6b)$$
The halo is usually assumed to be oblate
with $\varepsilon \sim 0.1-0.2$ 
(Nelson and Tremaine 1995).  This gives
$\kappa_{hz}^2-\Omega_h^2=(q^{-2}-1)\Omega_h^2 
\sim (0.23 - 0.56)\Omega^2_h$
which is of interest in the following.  
 For the
radii of the warps observed in 
galaxies ($r ~{\buildrel > \over \sim}
~4r_d$), $\kappa_{idz}^2-\Omega_{id}^2 \ll \Omega_h^2$.
For $q=1$, the halo mass within a radius $r$ is
$M_h(r)\approx 0.925\times10^{11}{\rm M}_\odot(r/10{\rm kpc})
(v_o/200{\rm km/s})^2 [r^2/(r^2+r_o^2)].$

Perturbations of the galaxy 
are assumed to consist 
of small angle tilting ($\theta^2\ll 1$)
of the outer disk with 
azimuthal mode number $m=1$ (or $-1$).  
The halo
and the inner disk are assumed to be unaffected by the 
perturbation.  We can describe the outer disk 
by a number $N$ 
of tilted plane circular rings.  
 This description is
{\it general} for small tilting angles where  
the linearized equations are applicable.
 We do not need to assume that the disk
is `razor thin' but rather that the disk 
half-thickness $\Delta z$ is small, $(\Delta z)^2 \ll r^2$.
The vertical displacements of the
disk $\sim \theta r$ are in general much
larger than $\Delta z$.

\subsection{One Tilted Ring}

The tilting of the ring is completely described by the 
tilt angle $\theta(t)$, which is the angle between the
upward normal
to the ring and the $z-$axis, and the azimuthal angle
$\varphi(t)$, which is the angle between the 
line-of-nodes (where the ring intersects
the $z=0$ plane) and the $x-$axis.  The geometry
is shown in Figure 3.  The perturbation is assumed to
be small in the respect that $\theta^2 \ll 1$.  The
position vector of a point on the ring is
$${\bf R}=\left[~r{\rm cos}\phi,~ r{\rm sin}\phi,~h(\phi,t)~\right]~,
~~0\leq \phi \leq 2\pi~,
\eqno(7)$$
(Cartesian components), where $h(\phi,t)$ is the 
height of the ring above
the $z=0$ plane.  We have
$$h(\phi,t)=r\theta(t){\rm sin}\left[\phi-\varphi(t)\right]~,$$
$$~~~~~~~~~~~~~~~~=r\theta_x(t){\rm sin}\phi-r\theta_y(t){\rm cos}\phi~,
\eqno(8a)$$
where 
$$\theta_x(t) \equiv \theta(t){\rm cos}\left[\varphi(t)\right]
\eqno(8b)$$
is the angle of the tilt about the $x-$axis, and
$$\theta_y(t) \equiv \theta(t){\rm sin}\left[\varphi(t)\right]
\eqno(8c)$$ is the
angle about the $y-$axis.  
We have $\theta = \sqrt{\theta_x^2+\theta_y^2}$
and $\varphi={\rm tan}^{-1}(\theta_y/\theta_x)$.

\placefigure{fig3}

The angular momentum of the tilted ring is
$$~~~~~{\bf L}=M\int_0^{2\pi} {{d\phi}\over{2\pi}}~
{\bf R}\times{{d{\bf R}}\over{dt}}~,$$
$$=M\Omega r^2{\hat{\bf z}} +{\bf \delta L }~~,
\eqno(9)$$
where ${\bf \delta L}={\bf \hat x}{\delta L_x}+
{\bf \hat y}\delta L_y $, $M$ is the ring mass,
and $\Omega$ is its angular rotation rate which is
unaffected by the perturbation to first order in
$\theta$.  We have
$${{d{\bf R}}\over {dt}}= {{\partial{\bf R}}\over
{\partial s}}\bigg|_t{{ds}\over{dt}} + {{\partial{\bf R}}
\over{\partial t}}\bigg|_s$$
 $$={\hat {\bf t}}\Omega r + 
{{\partial{\bf R}}\over{\partial t}}\bigg|_s~.
\eqno(10a)$$
 Here, ${\hat{\bf t}}=(1/r)(\partial {\bf R}/\partial \phi)=
[-{\rm sin}\phi,~{\rm cos}\phi,~\theta_x {\rm cos}\phi$ $ +\theta_y
{\rm sin}\phi]$ is the unit tangent
vector to the ring (to first order in $\theta$), 
$s$ is the distance along the ring circumference measured
from, say, the ascending node where $h(\phi,t)=0$, and
$${{\partial{\bf R}}\over{\partial t}}\bigg|_s
=\left[0,~0,~r{{d\theta_x}\over{dt}}{\rm sin}\phi-r{{d\theta_y}\over{dt}}
{\rm cos}\phi\right]~.
\eqno(10b) $$
 From equation (9) we have
$${ \delta L}_x =
Mr^2\left(~\Omega \theta_y~ +{1\over2}
{{d\theta_x}\over{dt}}\right)~,
\eqno(11a)$$
$${ \delta L}_y =Mr^2\left(-\Omega \theta_x +{1\over2}
{{d\theta_y}\over{dt}}\right)~.
\eqno(11b)$$

The torque on the ring is
$${\bf T} = \int_0^{2\pi}rd\phi~{\bf R \times F}~,
\eqno(12a)$$
or $$T_x=~~\int_0^{2\pi}rd\phi~\left(r{\rm sin}\phi F_z - hF_y\right)
~,
\eqno(12b)$$ 
$$T_y=-\int_0^{2\pi}rd\phi~\left(r{\rm cos}\phi F_z - hF_x\right)
~. 
\eqno(12c)$$
 The ring is assumed to exert no torque on itself.  
 Consequently, the
force ${\bf F}$ (per unit circumference) 
which contributes to ${\bf T}$ in
equation (12a)
is due to the inner 
disk and the halo. 
Thus we have 
$$F_z=-(M/2\pi r){\overline \kappa}_z^2 z~,
$$ 
where 
$${\overline \kappa}^2_z \equiv
\kappa_{idz}^2+\kappa_{hz}^2~,
\eqno(13a)$$
is the vertical epicyclic frequency excluding
the contribution due to the outer disk 
[see equation (3)].
Also, 
$$F_x=-(M/2\pi r){\overline \Omega}^2 r{\rm cos}\phi~,
~ F_y= -(M/2\pi r){\overline \Omega}^2r{\rm sin}\phi~,
$$
where
$${\overline \Omega}^2 \equiv \Omega_{id}^2+\Omega_h^2~,
\eqno(13b)$$
which excludes the outer disk contribution to $\Omega^2$ 
[see equation (2)].  
Thus we find
$$T_x=-{1\over2}M r^2({\overline \kappa}_z^2-
{\overline \Omega}^2)\theta_x~,
\eqno(14a)$$
$$T_y=-{1\over2}M r^2({\overline \kappa}_z^2-
{\overline \Omega}^2)\theta_y~.
\eqno(14b)$$

The equations of motion $d({\bf \delta L})/dt = {\bf T}$ are
$${d\over dt}\left(I{d\theta_x\over dt}+2I\Omega\theta_y\right)
=-I\left({\overline \kappa}_z^2-
{\overline \Omega}^2\right)\theta_x~,
\eqno(15a)$$
$${d\over dt}\left(I{d\theta_y\over dt}-2I\Omega\theta_x\right)=
-I\left({\overline \kappa}_z^2-
{\overline \Omega}^2\right)\theta_y~,
\eqno(15b)$$
after multiplying by two and letting $I=Mr^2$ denote the
moment of inertia of the ring.
The terms of the left-hand side $\propto \Omega$ are due
to the Coriolis force.

It is useful to introduce
$$\Theta(t)\equiv\theta_x(t)-i\theta_y(t)=\theta(t)
{\rm exp}\left[-i\varphi\left(t\right)\right]~.
\eqno(16)$$
This representation is well-known
from treatments of spin precession in
quantum mechanics.
We can then combine equations (15a) and (15b) to obtain
$$\left({d^2\over dt^2}+2i\Omega{d\over dt}\right)\Theta 
= - \Delta^2\Theta~,
\eqno(17)$$
where 
$$\Delta^2\equiv {\overline \kappa}^2_z-{\overline \Omega}^2~.
\eqno(18)$$
Note that we can have $\Delta^2<0$ for a prolate halo.
Solutions of equation (17) can be taken as
$\Theta =C{\rm exp}\left(-i\omega t\right)$ 
with $C$ a complex
constant.  
From equation (17) we have the dispersion relation
$~(\omega-\Omega)^2=\Omega^2+\Delta^2$ which gives
$$\omega=\Omega \pm \sqrt{\Omega^2+\Delta^2}~.
\eqno(19)$$
In all cases considered here, $\Omega^2+\Delta^2>0$.
For $\Omega>0$, the plus sign 
corresponds to the {\it fast
mode}, $\omega_2>0$, with 
prograde precession [ $\varphi(t)$ increasing ], 
and the minus sign to the {\it slow
mode}, $\omega_1$,  with retrograde 
precession [ $\varphi(t)$ decreasing ] (HT).  
For $|\Delta^2|\ll \Omega^2$
and $\Omega>0$,
the slow mode has $\omega_1 \approx -\Delta^2/(2\Omega)$.  
 These
two modes are the analogues 
of the {\it normal modes} of
vibration of a non-rotating mechanical system.
 That there are two modes rather than one
for each ring is due to the Doppler splitting
from the rotation ($\Omega$).

A general solution of equation (17) is
$$\Theta(t)=C_1{\rm exp}(-i\omega_1 t)+
C_2{\rm exp}(-i\omega_2 t)~,
\eqno(20)$$
where $\omega_1$ and
$\omega_2$ are the two roots given in equation (19), and  
$C_1$ and $C_2$ are complex constants.  
 The four real quantities
determine the initial values 
$\theta_x(0)$, $\theta_y(0)$, $d\theta_x(0)/dt$, and
$d\theta_y(0)/dt$.  
 If only the fast or slow mode is 
excited ($C_1$ or $C_2=0$), then $\theta$ = Const.
and $\varphi=\omega_\alpha t+$ Const.$^\prime$ 
with $\alpha = 1,2$.  
 A simple way to excite
only the slow or fast mode is to take $\theta_x(0)=C_\alpha$,
$\theta_y(0)=0$, $d\theta_x(0)/dt=0$, and $d\theta_y(0)/dt=
\omega_\alpha C_\alpha$.  Figure 4 illustrates the two limiting
cases of $C_1/C_2 \ll 1$ and $C_1/C_2 \gg 1$.

We can recast equations (15) in a 
form equivalent to that given by HT
by multiplying equation (15a) by 
${\rm sin}\phi$ and equation (15b) by
${\rm cos}\phi$ and subtracting the two.  
Noting that $\partial h/\partial \phi
=r\theta_x{\rm cos}\phi + r\theta_y{\rm sin}\phi$ 
and $\partial^2h/\partial \phi^2 =
-h$, we obtain
$$\left({\partial\over{\partial t}}+
\Omega{\partial\over{\partial \phi}}\right)^2
h = -\left({\overline \kappa}_z^2+
\Omega^2-{\overline \Omega}^2\right)h~.
$$
Note that $h(r,\phi,t)$ is non-zero
only for $r$ equal to the ring radius.
From equation (2) we have
$$\Omega^2-{\overline\Omega}^2=
{1\over r}{{\partial\Phi_{ring}}\over{\partial r}}~.
$$
For a toroidal ring of minor radius 
$\Delta r \ll r$, this difference is of the
order of $GM\ell n(r/\Delta r)/r^3$.  
In the absence of a halo, the right
hand side of the equation for $h$
is 
$$-(\kappa_{idz}^2+\Omega^2-{\overline \Omega}^2)h~,
$$
which is the same as the HT expression
$$-G\int d^2r'~ {{\Sigma(r')
\left[h(r,\phi,t)-h(r',\phi',t)\right]}\over
{|{\bf r - r'}|^3}}~.
$$
The $\kappa_{idz}^2$ term comes 
from the integration of over $r'<r$ while
the $\Omega^2-{\overline \Omega}^2$ 
term is from the integration over $r' \approx r$.

Multiplying equation (15a) by 
$d\theta_x/dt$ and (15b) by $d\theta_y/dt$
and adding the two gives $d{\cal E}_{ring}/dt = 0$, where
the energy of the tilted ring is
$${\cal E}_{ring} = 
{1\over 2}I\left({\dot \theta}_x^2+\dot\theta_y^2 \right)^2
+{1\over2}I\Delta^2\left(\theta_x^2+\theta_y^2 \right) = {\rm Const.~,}
\eqno(21)$$
where a dot denotes a time derivative. 
This expression agrees with 
the result of HT (their Appendix B).  In
the absence of a halo or with 
an oblate halo ($\kappa_{hz}^2 >\Omega_h^2$),
we have $\Delta^2 >0$, so 
that both terms in ${\cal E}_{ring}$ are
non-negative which implies 
stability of the ring tilting.  In
the case of a prolate halo 
with $\Delta^2<0$ instability is possible.

 In order to understand the
stability of the ring tilting, it is 
useful to consider the influence of a small
Newtonian drag or friction on 
the ring motion, a force ${\bf F}^f =
-(M/2\pi r)\beta \partial {\bf R}/\partial t|_s$ 
per unit
circumference, 
with $\beta >0$ but $\beta^2 \ll \Omega^2$.  
 This drag
could result from dynamical friction
of the ring with the dark halo matter 
(Nelson and Tremaine 1995).

 The full calculation of $\beta$ due to
dynamical friction on a precessing ring is
complicated (Nelson and Tremaine 1995), 
but an estimate based on treating
the ring as two point masses $M/2$ at the
points ${\rm max}|h|$ (Weinberg 1985) gives
$$\beta/\Omega \sim 2fGM\ell n\Lambda/
(3\sqrt{\pi} v_o^2 r)$$ 
$$\approx 0.035 \ell n
\Lambda \left({M \over 10^{10} {M}_\odot}\right)
\left({200{\rm km/s}\over v_o}\right)^2
\left({10{\rm kpc}\over r}\right)~,
\eqno(22)$$ 
where $\ell 
n\Lambda = {\cal O}(3)$ is the
Coulomb logarithm, $v_o$ is the circular velocity,
and $f=r^2(r^2+3r_o^2)/(r^2+r_o^2)^2$ accounts
for the density profile of the halo, with $r_o$ the halo
core radius.  
 Later, in \S 3, where we represent the outer
disk as $N$ tilted interacting rings, the
pertinent estimate for $\beta_j$ for
a single ring is this formula with $M$ 
the total mass of the outer disk rather than
the mass of a single ring (Weinberg 1985).

 The drag torque due to ${\bf F}^f$ is
$T_x^f = - {1\over2}Mr^2\beta$ $(d\theta_x/dt)$ and
$T_y^f = - {1\over2}Mr^2\beta(d\theta_y/dt)$.  
Solution of
equation (17) including ${\bf T}^f$ gives  
$$\omega = \Omega \pm \sqrt{\Omega^2+\Delta^2}-
{i\beta\over2}\left(1\pm{\Omega\over
\sqrt{\Omega^2+\Delta^2}}\right)~.
\eqno(23)$$
 Thus, both fast and slow precession 
modes with $\Delta^2>0$
are stable in the presence of friction, 
whereas the slow precession mode
with $\Delta^2 <0$ is unstable 
$~[~\omega_i\equiv Im(\omega)>0~]$.
 For the slow precession mode, the kinetic energy
term $\propto \dot \theta_{x,y}^2$ in ${\cal E}_{ring}$
is smaller than the 
magnitude of the potential energy term
$\propto \theta_{x,y}^2$ for $|\Delta^2| < \Omega^2$.
In the presence of the friction force,
$d{\cal E}_{rings}/dt = -I\beta(\dot\theta_x^2+\dot\theta_y^2)$.
 Thus, for the slow precession mode,
${\cal E}_{ring}\propto \Delta^2$ is negative 
if $\Delta^2 < 0$, and the tilt angle grows.  
 The instability is therefore a {\it negative
energy dissipative} instability.
An analogous instability was 
predicted and observed for 
the precession of laboratory 
collisionless relativistic 
electron rings (Furth 1965; Beal et al. 1969).  The
instability also occurs in the presence of
a non-Newtonian (nonlinear) friction force
on the ring.

A second quadratic integral of the motion of
equations (15), analogous to 
angular momemtum (see following), 
can be obtained by multiplying
equation (15a) by $-\theta_y$ 
and equation (15b)
by $\theta_x$ and adding.  
 This gives
$d{\cal P}_{ring}/dt = 0$, where
$${\cal P}_{ring} =I\left(\theta_y{\dot \theta_x}-
\theta_x{\dot \theta_y}
\right)+I\Omega\left(\theta_x^2
+\theta_y^2\right)~,
$$
$$=I \theta^2 \left(\Omega-\dot\varphi\right)~= {\rm Const.}
\eqno(24)$$
An analogous constant of the motion exists for cases of
two or more rings (see \S\S 2.3, 3, and 4).  
 This constant of the motion
is a new result of the present work. 
 For the case of instability, with
$\omega$ complex, $d\phi/dt = \omega_r \equiv Re(\omega)$.
With the above-mentioned friction force included, we
have $d{\cal P}_{ring}/dt = 
\beta I \theta^2\omega_r$, which
is compatible with equation (23) which shows instabililty
for $\Delta^2<0$ and gives $0<\omega_r<\Omega$ 
and ${\cal P}_{rings}>0$.

The second integral (24) can be understood by
noting that equations (15) follow from the 
Lagrangian
$${\cal L}={1\over 2}I\left(\dot\theta_x^2+\dot\theta_y^2\right)+
I\Omega\left(\theta_y\dot\theta_x-\theta_x\dot\theta_y\right)$$
$$
-{1\over 2}I\Delta^2\left(\theta_x^2+\theta_y^2\right)~.
\eqno(25a)$$
A canonical transformation to the variables $\theta$ and $\varphi$
gives
$${\cal L}={1\over 2}I\left(\dot \theta^2+\theta^2\dot\varphi^2\right)
-I\Omega\theta^2\dot\varphi-{1\over 2}I\Delta^2 \theta^2~.
\eqno(25b)$$
Thus the canonical momentum $P_\varphi 
=\partial {\cal L}/\partial \dot \varphi$
$=I\theta^2$ $(\dot\varphi-\Omega)$ = Const. $= -{\cal P}_{rings}$ 
because $\partial {\cal L}/\partial \varphi
=0$. On the other hand, $P_\theta=
I\dot \theta$ depends on time.  
The Hamiltonian ${\cal H}(\theta,\varphi,P_\theta,P_\varphi)=
\dot \theta(\partial {\cal L}/\partial \dot \theta)+\dot
\varphi(\partial {\cal L}/\partial \dot \varphi)-{\cal L}$ is
$${\cal H}={P_\theta^2\over 2I}+
{1\over 2I}\left({P_\varphi\over\theta}
+I\Omega\theta\right)^2+{1\over 2}I\Delta^2\theta^2~,
\eqno(25c)$$
and ${\cal H}={\cal E}_{ring}$ = Const. in that ${\cal L}$ 
has no explicit time 
dependence.  

The $\theta-$dependent terms of ${\cal H}$ can
evidently be viewed as an effective potential for the
$\theta$ motion. In general (for $P_\varphi \neq 0$), $\theta$
nutates between a minimum and a maximum value.  
The angular frequency of the nutation 
is $2(\Omega^2+\Delta^2)^{1\over2}$
so that the {\it nutation period} is 
$P_{nut} =(P/2)(1+\Delta^2/\Omega^2)^{-{1\over2}}$,
where $P=2\pi/\Omega$ is the period of the orbit.  
This can also be seen by noting that
with $\Theta(t)$ given in general by equation (20), we
have $|\Theta(t)|=|C_1+C_2{\rm exp}[-i(\omega_2-\omega_1)t]|$ and
$\omega_2-\omega_1 = 2(\Omega^2+\Delta^2)^{1\over2}$.
 For $|\Delta^2|/\Omega^2 \equiv \epsilon \ll 1$, the
period of the slow mode $P_1=2\pi/|\omega_1| \approx
(2/\epsilon)P\gg P_{nut}$, while the fast mode period
$P_2=2\pi/\omega_2 \approx(P/2)(1-\epsilon/4)$ is slightly
longer than the nutation period 
$P_{nut}\approx(P/2)(1-\epsilon/2)$.
 The
nutation of the ring is clearly evident in Figure 4.
 For $|C_1/C_2|<1$ (Figure 4a), $\Theta(t)$ nutates many 
(~$\approx 4/\epsilon~$) times in the period $P_1$ of
motion of $\Theta$ about the origin. 
On the other hand, for $|C_1/C_2| >1$ (Figure 4b), $\Theta(t)$
has an elliptical path with 
the azimuth of say the
maximum of $|\Theta(t)|$ precessing slowly in the clockwise
direction with a period $\approx (2/\epsilon)P$.  

\placefigure{fig4}

A counter-rotating ring with 
$\Omega \rightarrow -\Omega$ behaves in
the same way as a co-rotating ring.  
For $\Delta^2>0$, 
the ring slow mode precession is retrograde 
relative to the ring particle 
motion but of course prograde 
relative to our coordinate system.

\subsection{Interpretation of Ring Observations}

Warps of spiral galaxies are 
deduced from measurements
of $21$ cm line neutral hydrogen emission.  
Here,
we consider the spectral 
signature of a single ring in either
the slow precession mode or 
the fast precession mode.
The observer is considered
to be in the $(x,z)$ plane at an angle $\psi$
to the $z-$axis.  The velocity of the ring
matter is given by equation (10).  Thus, the
velocity of the HI in the direction of the
observer is
$$v_\parallel(\phi)/(r\Omega)= 
-~{\rm sin}(\psi){\rm sin}(\phi)~~~~~~~~~~~~~~~~~~$$
$$~~~~~~~~~~~~+~\theta
\left(1-{\omega_\alpha \over \Omega}\right)
{\rm cos}(\psi){\rm cos}(\phi-\varphi)~.
\eqno(26)$$
The spectrum of the ring is ${\cal S}(v_\parallel) 
\propto |dv_\parallel/d\phi|^{-1}$ $\propto
(1-v_\parallel^2/v_{\parallel~max}^2)^{-{1\over 2}}$.
 The ring motion $\partial {\bf R}/\partial t|_s$ enters
through the term involving $\omega_\alpha$.  
 For a ring
in the slow precession mode with $|\omega_1/\Omega| \ll 1$,
the ring motion has only a small affect on $v_\parallel(\phi)$.
 The situation is very different for a ring in the fast
precession mode where $\omega_2/\Omega > 2$ if
$\Delta^2 >0$.  
 In this case the sign of the term
in $v_\parallel(\phi)$ proportional to $\theta$
is reversed which corresponds to a change of the 
line-of-nodes angle 
$\varphi$ by $\pm 180^o$.  Observations
are interpreted assuming negligible ring motion,
$|\omega_\alpha/\Omega| \ll 1$, and this is wrong
if the fast precession mode(s) is excited.
  The fast precession modes of a system with many
rings may
disappear from view over a long enough time
 due to phase-mixing (see \S 3.1 and \S 3.2).

\subsection{Two Tilted Rings}

Consider now the case of two tilted plane circular rings of mass
$M_j$ and angular rotation rate $\Omega_j$ at radii $r_j$ with
$j=1,2$.  We have
$$\delta L_{jx}= 
 +{1 \over 2}I_j
{{d\theta_{jx}}\over{dt}}+I_j\Omega_j \theta_{jy}~,
\eqno(27a)$$
$$
~~\delta L_{jy}= +{1\over2}I_j
{{d\theta_{jy}}\over{dt}}~-I_j\Omega_j \theta_{jx}~,
\eqno(27b)$$
where $\theta_{jx}$ and $\theta_{jy}$ are the tilt angles and
$I_j\equiv M_jr_j^2$ the moments of inertia for the
two rings.

To obtain the torque on, say,
ring 1, we exclude as before the force due to this ring on itself.
Thus the horizontal force on ring $1$ is 
$$({\bf F}_1)_{x,y}=-M_1\left(\Omega_{id}^2+\Omega_h^2+{1\over r}
{{\partial \Phi_{R2}}\over{\partial r}}\right)_1\left(x_1,~y_1\right)~,
\eqno(28)$$
where the $\Phi_{R2}$ is the 
gravitational potential due to ring 2,
and the $1$-subscript on the parenthesis 
indicates evaluation at $r=r_1$.  Note
that
$$ \left({1\over r}{{\partial \Phi_{R2}}\over{\partial r}}\right)_1=
{1\over r_1}{\partial \over{~\partial r_1}}
\int_0^{2\pi}{{d\phi}\over{2\pi}}{GM_2\over{|{\bf r}_1-{\bf r}_2|}}~$$
$$~~~~~~~~~~~~~~~~~~~~~~=GM_2 I_{12}-GM_2 r_2 J_{12}/ r_1~,
\eqno(29)$$
where 
$$I_{12}\equiv \int_0^{2\pi}{{d\phi}\over{2\pi}}{1\over{\left(
r_1^2+r_2^2-2r_1r_2{\rm cos}\phi\right)^{3\over2}}}~,
\eqno(30a)$$
$$J_{12}\equiv \int_0^{2\pi}{{d\phi}\over{2\pi}}{{{\rm cos}\phi}\over{\left(
r_1^2+r_2^2-2r_1r_2{\rm cos}\phi\right)^{3\over2}}}~.
\eqno(30b)$$
There is also the vertical force on ring $1$,
$$F_{1z}=-\left(M_1/2\pi r_1\right)
\left(\kappa_{idz}^2+\kappa_{hz}^2\right)_1 h_1 +F_z^{12}~,
\eqno(31a)$$
where 
$$F_z^{12}=-\int_0^{2\pi}{d\phi'\over 2\pi}~
{GM_1M_2\left[h_1(\phi,t)-h_2(\phi',t)\right] \over 
\left[r_1^2+r_2^2-2r_1r_2{\rm cos}(\phi-\phi')\right]^{3\over2}}
\eqno(31b)$$ 
is the vertical force on ring 1 due to ring 2.  This
integral can be simplified by noting that
$$h_2(\phi',t)=r_2\theta_2 {\rm sin}(\phi-\varphi_2+\phi'-\phi)~$$
$$ = h_2{\rm cos}(\phi'-\phi)-r_2\theta_{2}{\rm cos}(\phi-\varphi_2)
{\rm sin}(\phi'-\phi)~.
\eqno(31c)$$
The sin$(\phi-\phi')$ term in equation (31c) 
does not contribute to the integral (31b).
Thus we have
$$\int_0^{2\pi}r_1d\phi~ {\bf R}_1\times \left({\hat{\bf z}}F_z^{12}
\right)=$$
$$-{1 \over 2}GM_1 M_2 r_1^2 I_{12} \theta_{1x}{\hat{\bf x}}+
{1 \over 2}GM_1 M_2 r_1 r_2 J_{12}\theta_{2x}{\hat {\bf x}}~
$$
$$-{1 \over 2} GM_1M_2r_1^2I_{12}\theta_{1y}{\hat {\bf y}}+
{1 \over 2}G M_1 M_2 r_1 r_2 J_{12} \theta_{2y}{\hat {\bf y}}~.
\eqno(32)$$
Combining terms in equations (12b) and (12c) gives
$$T_{1x}=-{1\over2}M_1 r_1^2\Delta_1^2\theta_{1x}-
{1\over 2}C_{12}\left(\theta_{1x}-\theta_{2x}\right)~,
\eqno(33a)$$
$$T_{1y}=-{1\over2}M_1 r_1^2 \Delta_1^2\theta_{1y}-
{1\over 2}C_{12}\left(\theta_{1y}-\theta_{2y}\right)~,
\eqno(33b)$$
where $$C_{12}\equiv GM_1 M_2 r_1 r_2 J_{12}
\eqno(34)$$ 
measures the 
strength of the gravitational 
interaction between the two rings, and
$\Delta^2$ is defined in equation (18).  
 The two terms
involving the integral $I_{12}$ cancel.
 Note that $C_{12}$ is symmetric
between the two rings and that it is
 non-negative because
$J_{12} \geq 0$ [ see equation (45e) and Figure 5 ].

The torque on ring 2 is found in the same way to be
$$T_{2x}=-{1\over2}M_2 r_2^2\Delta_2^2 \theta_{2x}-
{1\over 2}C_{12}\left(\theta_{2x}-\theta_{1x}\right)~,
\eqno(35a)$$
$$T_{2y}=-{1\over2}M_2 r_2^2 \Delta_2^2 \theta_{2y}-
{1\over 2}C_{12}\left(\theta_{2y}-\theta_{1y}\right)~.
\eqno(35b)$$
The equations of motion are 
$$d({\bf \delta L}_1)/dt ={\bf T}_1~,~~~ 
d({\bf \delta L}_2)/dt={\bf T}_2~.
\eqno(36)$$
For the limit where the only torques are those due to the two rings,
$\Delta^2_1 =0$ and $\Delta^2_2 =0$, we find 
$${d\over dt}\left( \delta{\bf L}_1 
+\delta{\bf L}_2\right)=0~,
\eqno(37)$$
which is a necessary result.

We can follow the steps leading 
to equation (22) to obtain
$d{\cal E}_{rings}/dt=0$ for 
the energy ${\cal E}_{rings}$ of the
two tilted rings, where
$${\cal E}_{rings}=~{1\over 2}\sum_{j=1}^2I_j
\left({\dot\theta_{jx}^2+\dot \theta_{jy}^2}\right)+
{1\over 2}\sum_{j=1}^2I_j\Delta_j^2\left( \theta_{jx}^2+\theta_{jy}^2\right)
$$
$$~~+{1\over 2}C_{12}\left(\theta_{1x}-\theta_{2x}\right)^2+
{1\over 2}C_{12}\left(\theta_{1y}-\theta_{2y}\right)^2~,
\eqno(38)$$
where $I_1 \equiv M_1r_1^2$ and 
$I_2\equiv M_2r_2^2$ are the moments
of inertia of the two rings.  
 This energy is non-negative and the rings are
stable if both $\Delta_1^2$
and $\Delta_2^2$ are non-negative. 
 Also, 
in this case the rings are stable
in the presence of dissipative 
forces such as the friction force
of Subsection 2.1.  
 Thus a necessary condition
for instability with or without
dissipation is that 
$\Delta_1^2$ and/or $\Delta_2^2$ be
negative.

A second quadratic integral of the motion
can be obtained by following the steps
leading to equation (24).  We find
$d{\cal P}_{rings}/dt = 0$, where
$${\cal P}_{rings} =
I_1\theta_1^2\left(
\Omega_1-{\dot\varphi_1}\right)+
I_2\theta_2^2\left(
\Omega_2-{\dot\varphi_2}\right)~.
\eqno(39)$$
For the case of instability, $\omega=\omega_r+i\omega_i$,
$\omega_i>0$, we have
$d\varphi_j/dt = \omega_r$.  
 Thus, a further necessary
condition for instability in the
absence of dissipation (in addition to $\Delta_1^2$
and/or $\Delta_2^2$ being negative) is that  
$\Omega_1-\omega_r$
and $\Omega_2-\omega_r$ have different signs; that is,
$\Omega_2<\omega_r<\Omega_1$ for $\Omega_2<\Omega_1$.
 For the
slow precession modes with $|\omega_r| \ll \Omega_j$,
this corresponds to counter-rotating rings, $\Omega_1/
\Omega_2 <0$.

The Lagrangian for the two ring system is
$${\cal L}={1\over 2}\sum_{j=1}^2I_j\left(\dot\theta_j^2
+\theta_j^2\dot\varphi_j^2
-2\Omega_j\theta_j^2\dot\varphi_j-\Delta_j^2\theta_j^2 \right)-$$
$${1\over 2}{C_{12}}\left[\theta_1^2+\theta_2^2-
2\theta_1\theta_2{\rm cos}(\varphi_1-\varphi_2)\right]~.
\eqno(40)$$
The corresponding Hamiltonian
${\cal H}=\sum (\dot \theta_j \partial 
{\cal L}/\partial \dot \theta_j 
+\dot \varphi_j \partial {\cal L}/\partial \dot\varphi_j)
-{\cal L}$ is equal in value to ${\cal E}_{rings}$
of equation (38).  
 The second constant of the motion, ${\cal P}_{rings}=
-P_{\varphi 1}-P_{\varphi 2}$, 
where $P_{\varphi j} \equiv \partial {\cal L}
/\partial \dot\varphi_j$.  
 Note that 
$d P_{\varphi 1}/dt=\partial {\cal L}/\partial \varphi_1
=-{1\over 2}\theta_1\theta_2{\rm sin}(\varphi_1-\varphi_2)$ 
and that
$d P_{\varphi 2}/dt=\partial {\cal L}/\partial \varphi_2
={1\over 2}\theta_1\theta_2{\rm sin}(\varphi_1-\varphi_2)$ so
that $d(P_{\varphi 1}+P_{\varphi 2})/dt = 0$.

As in the case of a single ring, 
it is useful
to introduce
$$\Theta_1(t)\equiv \theta_{1x}-i\theta_{1y}=\theta_1(t){\rm exp}
\left[-i\varphi_1(t)\right]~,
\eqno(41a)$$
$$\Theta_2(t)\equiv \theta_{2x}-i\theta_{2y}=\theta_2(t){\rm exp}
\left[-i\varphi_2(t)\right]~.
\eqno(41b)$$
The equations of motion (36) can then be written as
$$\left({d^2 \over dt^2}+2i\Omega_1{d\over dt}\right)\Theta_1=
-\Delta_1^2\Theta_1-{ C_{12}\over I_1}\left(\Theta_1 -\Theta_2\right)~,
\eqno(42a)$$
$$\left({d^2\over dt^2}+2i\Omega_2{d\over dt}\right)\Theta_2=
-\Delta_2^2\Theta_2-{ C_{12}\over I_2}\left(\Theta_2 -\Theta_1\right)~,
\eqno(42b)$$
after dividing through by $I_j/2$.

For $\Theta_1$ and $\Theta_2 \propto
{\rm exp}(-i\omega t)$, we get the 
dispersion relation
$$\left[\left(\omega-\Omega_1\right)^2 -K_1^2\right]
\left[\left(\omega-\Omega_2\right)^2-K_2^2\right]={C_{12}^2
\over I_1I_2}~,
\eqno(43)$$
where
$$K_j^2 \equiv \Omega_j^2+ \Delta_j^2+{C_{12}\over I_j}~,
$$
with $j=1,2$.  Equation (43) can readily be solved for the
four roots.  Two of these roots are high-frequency, {\it fast
precession modes} with $\omega \geq 2\Omega_j$, assuming
$\Omega_j >0$ and $\Delta_j^2 \geq 0$.  The other two
roots are the {\it slow precession modes} with much smaller
frequencies, typically, $\omega^2 \ll \Omega_j^2$.

In order to get simple analytic results, we first
consider the low-frequency modes $\omega^2 \ll \Omega_j^2$, 
assuming $|\Delta_j^2| \ll \Omega_j^2$ 
and $C_{12}/I_j \ll \Omega_j^2$.  We can then neglect
the second time derivatives in equations (42) and the
$\omega^2$ terms in equation (43), and this leads to
a quadratic dispersion equation for $\omega$.  The
roots of this equation are
$$\omega ={1\over2}\left[\omega_{p1}+\omega_{p2}-\omega_{g12}\left(
1+\ell_{12}\right)\right]
$$
$$~~
\pm ~{1\over2}\left\{\left[\omega_{p2}-\omega_{p1}+\omega_{g12}\left(1-
\ell_{12}\right)\right]^2+4\omega_{g12}^2\ell_{12}\right\}^{1\over2}~.
\eqno(44)$$
Here,
$$\omega_{pj} \equiv -~{\Delta_j^2 \over 2\Omega_j}~,
\eqno(45a)$$
for $j=1,2$, are the slow precession frequencies of the two
rings in the absence of gravitational 
interaction between them ($C_{12}=0$); 
$$\omega_{g12} \equiv {C_{12}\over 2 L_1} ~~{\rm and }~~~
\ell_{12} \equiv {L_1 \over L_2}~,
\eqno(45b)$$
where $\omega_{g12}$ is a frequency which 
measures the strength of the gravitational interaction
between the two rings, and
where $L_j \equiv I_j\Omega_j$ are the angular momenta of
the two rings. 
We have
$${\omega_{g12} \over \Omega_1}\approx
2.8\times 10^{-2}\left({M_2 \over 10^{10}~{\rm M_\odot}}\right)\times
$$
$$\left( {20~{\rm kpc}\over r_2}\right)\left({ 200~{\rm km/s}\over
\Omega_1 r_1}\right)^2 J^\prime_{12}~,
\eqno(45c)
$$
where
$$J_{12}^\prime\equiv \delta 
\int_0^{2\pi}{d\phi\over 2\pi}~{{\rm cos}\phi\over
\left[1+\delta^2-2\delta {\rm cos}\phi\right]^{3\over2}}~~,
\eqno(45e)$$
with $\delta = r_1/r_2 \leq 1.$    
For $\delta \ll 1$, 
$J_{12}^\prime \rightarrow {3\over 2}\delta^2$.
For $1-\delta \ll 1$, $J_{12}^\prime \rightarrow 1/[\pi(1-\delta)^2]$.  
We
assume that $1-\delta > \Delta z/r$, where $\Delta z$ is the disk 
thickness.  The dependence of $J_{12}^\prime$ on $\delta$ is shown
in Figure 5. 

\placefigure{fig5}

A simple limit of equation (44) 
is that where $\Delta_1^2=0=\Delta_2^2$.
The two roots are then $\omega=0$ 
and $\omega= -\omega_{g12}(1+\ell_{12}) =
-{1\over 2}C_{12}L_1L_2/(L_1+L_2)$. 
The zero frequency mode corresponds
to both rings tilted by the 
same angle, $\Theta_1=\Theta_2$, which is
the rigid tilt mode (HT).  
The other mode has $\Theta_1=-(L_2/L_1)\Theta_2$;
that is, the rings are tilted 
in opposite directions for $L_1/L_2 >0$.

We first consider   
co-rotating rings
where the four $\omega$ 
roots of equation (43) are 
real in the absence of
dissipation.  Note that
the low frequency roots of
equation (44) are real for $\ell_{12} >0$.
A general solution for the 
motion of the two rings is then
$$\Theta_j = \sum_{\alpha=1}^4 C_{j\alpha}~ 
{\rm exp}(-i\omega_\alpha t)~,
\eqno(46)$$ 
where $\{\omega_\alpha\}$ are 
the four mentioned frequencies, and
$\{C_{j\alpha}\}$ are $8$ complex 
constants.  However, only $8$
 real quantities are needed to 
specify general initial conditions
because the $\{ C_{1\alpha}\}$ 
are related to the $\{C_{2\alpha}\}$
by equations (42).  If only a 
{\it single} mode of the system is 
excited, then $\Theta_1 = 
C_{1\beta}{\rm exp}(-i\omega_\beta t)$
and $\Theta_2 = C_{2\beta}{\rm exp}(-i\omega_\beta t) $.  
Equations (42) then imply that $\Theta_2/\Theta_1$ is real.
Because $\Theta_j = \theta_j{\rm exp}(-i\varphi_j)$, we conclude
that $\Theta_2/\Theta_1 \propto 
{\rm exp}[-i(\varphi_2-\varphi_1)]$
is real so that $\varphi_2-\varphi_1$ 
is either $0$ or $\pi$.  That
is, the rings are either tilted 
in the same direction and precess
together or they are tilted in 
opposite directions and also precess
together.  From equations (42),  $\Theta_2/\Theta_1=$ 
$I_1\left[K_1^2-(\omega_\beta-\Omega_1)^2\right]/C_{12}$.

Figure 6a shows the $\omega_{g12}$ 
dependence of the frequencies of
the two slow-precession modes 
obtained from equations (43) and
(44) for the case where both
rings are co-rotating and the
halo is {\it oblate}.  
The `slower' of the two modes has $\Theta_2/\Theta_1 >0$,
while the other has $\Theta_2/\Theta_1  <0$ as shown
in Figure 6b.  
The larger gravitational
torque in the second case 
accounts for the faster retrograde
precession. 
As $\omega_{g12}$ increases, 
the frequency of the
slower mode approaches
$$ \omega_1 \rightarrow 
{L_1\omega_{p1} \over L_1+L_2} 
+{L_2\omega_{p2}\over L_1+L_2}~.
\eqno(47)$$
and $\Theta_2/\Theta_1 \rightarrow 1$
for $\omega_{g12} \gg 
|\omega_{p2}-\omega_{p1}|/(1+\ell_{12})$,
where the $\omega_{pj}$ are defined in
equation (45a).  
 This
result can be derived from 
equation (43) or (44).  
 For the case of $N$ rings and 
sufficiently large $\omega_{gjk}$,
we find in general that the lowest frequency 
approaches the angular momentum weighted
average of the single ring precession
frequencies $\omega_{pj}$ [equation (45a)].
 In contrast with the dependence of $\omega_1$,
$\omega_2$ decreases monotonically 
as $\omega_{g12}$ increases.

\placefigure{fig6}

From the approximate equation (44) for the
slow precession modes, we can distinguish
{\it weak} and {\it strong}-{\it coupling}
limits of two co-rotating rings.  The
coupling strength is measured by the 
dimensionless parameter
$$\xi_{12} \equiv { \omega_{g12}\left(1+\ell_{12}\right)
\over |\omega_{p2}-\omega_{p1}|}
$$
$$={GJ_{12}\left(L_1+L_2\right)\over
2(\Omega_1 r_1)(\Omega_2 r_2)|\omega_{p2}-\omega_{p1}|}~,
\eqno(48)$$
which is symmetrical in the ring indices (unlike
$\omega_{g12}$).
 For {\it weak coupling}, $\xi_{12} \ll 1$, the two
rings are affected little by their
gravitational interaction, and the two
slow mode frequencies are $\omega_1 \approx \omega_{p2}$
with $\Theta_2/\Theta_1 \gg 1$ and $\omega_2 \approx
\omega_{p1}$ with $|\Theta_1/\Theta_2| \gg 1$.
 In the opposite limit of {\it strong coupling},
$\xi_{12} \gg 1$, the gravitational interaction of
the two rings is important, and $\omega_1$ is
given approximately by equation (47) with $\Theta_2/\Theta_1
\sim 1$, and $|\omega_2| \gg |\omega_1|$ with
$\Theta_2/\Theta_1 <0$. 
 
 For nearby rings,
$(\Delta r)^2 = (r_2-r_1)^2 \ll r_2^2$, of equal mass
$M_1=M_2 = 2\pi r \Delta r~ \Sigma$ (with $\Sigma$ the
surface mass density of the disk at $r$), 
we have $J_{12} \approx 1/[\pi r (\Delta r)^2]$,
$~|\omega_{p2}-\omega_{p1}| 
\approx |\partial \omega_p/\partial r|\Delta r$, and thus
$$\xi_{12} \approx {2 G \Sigma 
\over \Omega (\Delta r)^2 \left|{\partial \omega_p /
\partial r}\right|}~.
\eqno(49)$$
 Evidently, there is strong (weak) coupling for
$\Delta r \ll \Delta r_c$ ($\Delta r \gg \Delta r_c$),
where $\Delta r_c$ corresponds to $\xi_{12} = 1$ so that
$$ \Delta r_c = \left[ {2 G \Sigma \over
\Omega |\partial \omega_p/\partial r|}\right]^{1\over2}~.
\eqno(50a)$$
Assuming a flat rotation curve ($\Omega r =$ const.) and
$(\Delta/\Omega)^2=$ const.,
we obtain
$$
\Delta r_c \approx 13.2~{\rm kpc}\left({200~{\rm km/s}\over \Omega r}
\right) \left({r\over 20~ {\rm kpc}}\right)^{3\over2} \times
$$
$$\left({0.2 \over |\Delta^2|/\Omega^2}\right)^{1\over 2}
\left({\Sigma \over 10~ {\rm M_\odot/pc^2}}\right)^{1\over2}~.
\eqno(50b)$$
Rings closer together than $\Delta r_c$ will be strongly
coupled.

\placefigure{fig7}

Figure 7 shows sample orbits of $\Theta_1$ and $\Theta_2$
for two co-rotating rings in an oblate halo for initial
conditions $\Theta_1/\Theta_2 = 1/2$ and $\dot \Theta_1=0=
\dot \Theta_2$.  The top panel shows a case of
weak coupling (small $\xi_{12}$) where 
the line-of-nodes $\varphi_j$ of the
two rings regress essentially independently with the
result that $\varphi_2-\varphi_1$ increases linearly 
with time.  The bottom panel is for a case of strong
coupling (large $\xi_{12}$) where the line-of-nodes
are {\it phase-locked} in the sense that the
time average of $\varphi_2-\varphi_1$ is zero.
Figure 8 shows the azimuthal angle difference
of the line-of-nodes for the two cases of 
Figure 7.  In the case of weak coupling,
a low frequency initial perturbation excites
both low frequency modes which have
similar frequencies ($\omega_2 \sim 2\omega_1$),
and this gives the unlocked behavior.  For
strong coupling, the initial perturbation
excites mainly the $\omega_1$ mode because
$|\omega_2| \gg |\omega_1|$, and this gives
the locked behavior.  
 For the parameters of figures 4 and 5, note that
$\xi_{12} \approx 60 \omega_{g12}/\Omega_1$. 

\placefigure{fig8}

Figure 9a shows the $\omega_{g12}$ 
dependence of the frequencies of
the two slow-precession modes 
obtained from equation (43) 
for the case where both
rings are co-rotating and the
halo is {\it prolate}.
 (We have adopted the convention
of ordering the mode frequencies
by their magnitudes for $\omega_{g12} 
\rightarrow 0$.) 
 Figure 9b
shows that the $\omega_2$ mode 
has $\Theta_2/\Theta_1 >0$.
 As in the case of an oblate halo,
for large $\omega_{g12}$ or $\xi_{12}$,
$\omega_2$ approaches the limit
given by equation (47) and 
$\Theta_2/\Theta_1 \rightarrow 1$.

\placefigure{fig9}

 As in the 
case of a single ring in a prolate halo
potential [see equation (23)], small
friction torques, with friction 
coefficients $\beta_1$ and
$\beta_2$, 
 on two co-rotating rings in a prolate halo
lead to a negative energy
dissipative instability.  
 We find that
the slow modes with prograde precession,
$\omega_\alpha >0$, are unstable in the
presence of dissipation.  
 That is,
the $\omega_1$ mode in Figure 9
is unstable (stable) for $\omega_{g12}
<$ $(>)$ $ 0.02$, whereas
the $\omega_2$ mode is unstable for
all $\omega_{g12}$.  
 The two
high frequency modes are prograde
but are damped by
friction for all $\omega_{g12}$.
 As in the case of an oblate halo (Figure 8),
we observe phase-locking for
sufficiently large values of $\xi_{12}$.

Frictional torques between two co-rotating
rings in a prolate halo also give
rise to a negative energy dissipative
instability.  Such torques could
arise from the viscous interaction
between adjacent gaseous rings.  These frictional
torques do not change the total
angular momentum of the two rings.
Therefore, these frictional torques, if
Newtonian (linear) in nature, are 
{\it necessarily} of the form that they contribute
terms 
$$-\left({I_2 \over I_1}\right)^{1\over2}\beta_{12}
\left(\dot\Theta_1-\dot\Theta_2\right)~,
\eqno(51a)$$
$$-\left({I_1 \over I_2}\right)^{1\over2}\beta_{12}
\left(\dot\Theta_2-\dot\Theta_1\right)~,
\eqno(51b)$$
on the right hand sides of 
equations (42a) and (42b), respectively,
where $\beta_{12}$ is the friction 
coefficient (with units of angular
frequency). 
 For a gas with kinematic shear viscosity $\nu$ (with units
of $cm^2/s$) the terms (51) arise from the
momentum flux density or stress $T_{rz}$ due
to the different vertical velocities of the two
rings as a function of $\phi$.  
 This gives
$\beta_{12} = \nu/(\Delta r)^2$, where
$\Delta r = r_2-r_1$.
 The relevant viscosity is that due to
turbulence in the gas $\nu = \nu_t$ owing
to the smallness of the microscopic viscosity.   
 An estimate of $\nu_t$ can be made following the
proposal of Shakura (1973) and Shakura and
Sunyaev (1973) that $\nu_t = \alpha c_s \Delta z$,
where $c_s$ is the sound speed,
$\Delta z$ is the half-thickness of the disk,
and $\alpha$ is a dimensionless constant
thought to be in the range $10^{-3}$ to $1$. 
 In Sections 3 and 4 we discuss further the
frictional torques between adjacent rings.

\subsection{Two Tilted, Counter-Rotating Rings }

Here, we consider the case of two tilted rings
which are rotating in opposite directions.  
This situation is pertinent to observed 
counter-rotating galaxies (see, for example,
Jore, Broeils, and Haynes 1996).
From equation (44), instability is possible
(in the absence of dissipation) only for $\ell_{12}
<0$ in agreement with our discussion of equation (39).  
Thus, consider
ring $2$ to be counter-rotating  
($\Omega_2<0$ and $L_2<0$), while ring $1$ is co-rotating.
The region of instability is bounded by the two curves
$$\omega_{p2}-\omega_{p1}={\Delta_1^2\over 2|\Omega_1|}+
{\Delta_2^2 \over 2|\Omega_2|} $$
$$= \left(-1\pm 2|\ell_{12}|^{1\over2}
-|\ell_{12}|\right)\omega_{g12}~,
\eqno(52)$$
on which the square root in
equation (44) is zero.  The curves
are shown in Figure 10.  For parameters in the
region between the two curves there is instability, 
which is seen to occur only for $\Delta_1^2$
and/or $\Delta_2^2$ negative.  For
fixed values of $\omega_{p1}$, $\omega_{p2}$, and
$\ell_{12}$, instability occurs in general ($|\ell_{12}|\neq 1$)
for $\omega_{g12}$ larger than a critical value
but smaller than a second larger critical value.  The maximum growth 
rate is for $(\omega_{p2}-\omega_{p1})/\omega_{g12} =
-1-|\ell_{12}|$ (so that $\omega_{p2} < \omega_{p1}$), 
and it is ${\rm max}(\omega_i) = 
\omega_{g12}|\ell_{12}|^{1\over2}$.  This is a {\it dynamical
instability} in that it does not depend on
dissipation as in the case 
of a single ring with $\Delta^2<0$.
For conditions of maximum growth, 
$\Theta_1/\Theta_2 = -i/|\ell_{12}|^{1\over2}$, so that
$\theta_1/\theta_2 = |L_2/L_1|^{1\over2}$ and $\varphi_1-
\varphi_2 = \pi/2$.  We can thus view the co-rotating ring
as `torquing up' the more slowly precessing counter-rotating
ring by the gravitational interaction or vica-versa.

\placefigure{fig10}

Figure 11a shows the $\omega_{g12}$ 
dependence of the frequencies of
the two slow-precession modes 
obtained from equation (43)  
for the case where ring $2~$ is
counter-rotating ($\Omega_2 <0$) and 
ring 1 is co-rotating ($\Omega_1>0$) and the
halo is {\it oblate}.  
 Figure 11b shows that
the mode $\omega_1$ with prograde precession
involves mainly motion of ring 2, whereas
the $\omega_2$ mode with retrograde 
precession involves mainly ring 1.  
 For the conditions studied ($\omega_{g12}
\leq 0.1\Omega_1$), we find no 
phase-locking. 

\placefigure{fig11}

\subsection{Three Tilted Rings}

Extension of the results of Subsection 2.2
to the case of three tilted rings is straightforward.
Figure 12a shows the dependence of the three slow
precession frequencies on the strength of the
gravitational interaction measured by $\omega_{g12}$
with $\omega_{g23}/\omega_{g12}=$ Const.  
For increasing $\omega_{g12}$, the slowest 
mode ($\omega_1$) {\it slowly} approaches the limiting
frequency given by the generalization of equation
(47) to three rings.
 The other two frequencies decrease monotonically
with increasing $\omega_{g12}$.   
Figure 12b shows
the geometrical nature of the three slow
precession modes.
Figure 13 shows sample orbits $\Theta_j(t)$ 
for three rings for cases of weak and strong
coupling. 
Figure 14 shows different cases of phase-locking
for a three ring system.

\placefigure{fig12}

\placefigure{fig13}

\placefigure{fig14}

\section{N Tilted Corotating Rings}

The generalization of equation (40) gives
the Lagrangian for an $N$ ring system as
$${\cal L}={1\over 2}\sum_{j=1}^NI_j\left(\dot\theta_j^2
+\theta_j^2\dot\varphi_j^2
-2\Omega_j\theta_j^2\dot\varphi_j-\Delta_j^2\theta_j^2\right)$$
$$
-~{1\over2}\sum_{j<k}^NC_{jk}\left[\theta_j^2+\theta_k^2-
2\theta_j\theta_k{\rm cos}(\varphi_j-\varphi_k)\right].
\eqno(53a)
$$
The Hamiltonian, 
$${\cal H}(\theta_j,\varphi_j,
P_{\theta j}, P_{\varphi j})=\sum_k( {\dot \theta_k}
{\partial{\cal L}\over \partial \dot\theta_k}+
\dot\varphi_k{\partial{\cal L}\over
\partial {\dot \varphi_k}})-{\cal L}
$$
is 
$${\cal H}=\sum_{j=1}^N\left[{P_{\theta j}^2 \over 2 I_j}+
{1\over 2 I_j}\left({P_{\varphi j} \over \theta_j}+\Omega_j
\theta_j
\right)^2 + {I_j\over 2}\Delta_j^2 \theta_j^2\right]
$$
$$
+~{1\over2}\sum_{j<k}^NC_{jk}\bigg[\theta_j^2+\theta_k^2-
2\theta_j\theta_k{\rm cos}(\varphi_j-\varphi_k)\bigg]~.
\eqno(53b)
$$
Because the $\{C_{ij}\}$ are non-negative, all of the
terms in ${\cal H}$ are non-negative if $\Delta_j^2 \geq 0$
which is the case for oblate halo.  Thus the $N-$ring
system with $\Delta_j^2>0$ is stable in the presence of
dissipative forces. 
The total canonical angular momentum is
$$
P_{\varphi}^{tot}=\sum_{j=1}^N {\partial {\cal L} \over
\partial \dot \varphi_j} = \sum_{j=1}^N I_j \theta_j^2(\dot 
\varphi_j - \Omega_j)~.
\eqno(53c)$$
In the absence of dissipation ${\cal E}_{rings} = {\cal H}$
and ${\cal P}_{rings} = - P_\varphi^{tot}$ are constants
of the motion. 

As in the case of one or two rings, it is useful to let
$$\Theta_j(t) =\theta_{jx}(t)-i\theta_{jy}(t)=
 \theta_j(t){\rm exp}[-i\varphi_j(t)]~.
\eqno(54)$$
The equations of motion can then be written as
$$I_j\left({d^2 \over dt^2}+2i\Omega_j{d\over dt}\right)\Theta_j=$$
$$
-\Delta_j^2I_j\Theta_j-\sum_{k \neq j}^N
{ C_{jk}}\left(\Theta_j -\Theta_k\right),
\eqno(55)$$
with $j=1..N$, where the $C_{jk}$ generalize $C_{12}$ of
equation (34).

As discussed in \S 2.1, a Newtonian drag
force on the ring motion due to dynamical
friction with the halo matter can be taken
into account by including the term
$$
-\beta_j I_j{d\Theta_j \over dt}
\eqno(56)$$
on the right hand side of equation (55).  The influence
of the relative friction discussed in \S 2.3 can
be accounted for by including the terms
$$
-\beta^\prime_{j+{1\over2}}I_{j+{1\over2}}(\dot \Theta_j
-\dot \Theta_{j+1})
 -\beta^\prime_{j-{1\over2}}I_{j-{1\over2}}(\dot \Theta_j
-\dot \Theta_{j-1})
\eqno(57)$$
on the right hand side of equation (55).   Here,
$I_{j \pm {1\over2}} \equiv (I_j I_{j \pm 1})^{1 \over 2}$
and
$\beta^\prime_{j\pm{1 \over 2}} \equiv \beta_{j~j \pm1}=
[\nu_t/(\Delta r)^2]_{j+{1 \over 2}}$ with $\beta'_{1\over 2} =0$
and $\beta'_{N+{1\over 2}}=0$.
 With the dissipation terms (56) and (57)
included one can readily show that
$d{\cal E}_{rings}/dt \leq 0$ (see also \S 4).
We have not found an analogous result for
$d{\cal P}_{rings}/dt$ in the presence
of dissipation.

We have developed and tested codes to solve equations
(55) (including the terms (56) and (57)) for $N=7,~13,~25,$ and
$49$ rings.  For the results presented here,
the rings are taken to be uniformly spaced in $r$
with ring one at $r_1=10~{\rm kpc}$ and ring $N$
at $r_N=40~{\rm kpc}$ so that
$r_j = 10 +(j-1)\Delta r$, with $\Delta r =30/(N-1)$
in units of kpc.  The ring masses $M_j=2\pi r_j \Delta r
\Sigma(r_j)$ and moments of inertia $I_j=M_jr_j^2$ are 
calculated with $\Sigma(r_j)$ the sum of 
the surface density of the inner exponential disk
and that of the neutral hydrogen given in \S 2.  
The $\Delta_j^2$
are calculated using equations (4), (6), and
(18).  The coupling coefficients $\{C_{jk}\}$ are
evaluated using equations (34) and (30b) and stored.
Equations (55) are solved as four first order
equations for $\theta_{xj},~\dot \theta_{xj},~\theta_{yj},$
and $\dot \theta_{yj}$.  At the same time, an additional
equation,
$$ {d\varphi_j \over dt}= { \theta_{xj}\dot \theta_{yj} -
\theta_{yj}\dot \theta_{xj}\over \theta_{xj}^2+\theta_{yj}^2}~,
\eqno(58)$$
is also solved to give $\varphi_j(t)$, which is the
line-of-nodes angle relative to the $x-$axis.

 Comparisons of the temporal responses of 
$25$ and $49$ ring sytems for
different initial conditions 
show generally good agreement 
of the warps $\Theta_j$ for the
time intervals studied.
This indicates that $N=49$ rings 
gives a valid representation of a
continuous disk. 
 However, in contrast with a continuous disk (HT),
an $N-$ring disk has $2N$ discrete modes with
separate frequencies - $N$ `low-frequency' modes
and $N$ `high-frequency' modes. 
 Figure 15 shows the power spectrum 
of $\theta_{x25}(t)$ for a $49$ ring disk-halo
system which is initially given a random
perturbation with all radial wavelengths
excited and no damping.
 The frequencies $\omega$ are normalized
by the angular velocity of the inner
ring $\Omega_1$.  
 Note that 
$P_1 \equiv 2\pi/\Omega_1 \approx 0.308$ Gyr
for our reference values.
 The frequency differences between adjacent
modes is $\delta \omega \sim \Omega_1/N$.
 For treatment of initial value problems
of disk warping for time intervals $\leq t_{max}$,
frequency differences $\delta \omega$ 
less than $1/t_{max}$ are irrelevant.
 It is sufficient to have 
$\delta \omega \sim 1/t_{max}$ which 
corresponds to $N \sim \Omega_1 t_{max}$
$= 2\pi(t_{max}/0.3{\rm Gyr})$ rings.
This suggests that a larger number of rings
may be needed for our cases 
with $t_{max} \geq 8$ Gyr 
(Figures 18 and 19).
 This will be investigated in
a future work.

 The lowest
frequency mode of the power spectrum, 
$\omega_1$ in Figure 15 (see also \S 2.3 and \S 2.5),
remains discrete (that is, isolated) 
as $N\rightarrow \infty$
because our disk model has a sharp outer edge
so that $1/\Sigma(r)$ is integrable (HT; Sparke
and Casertano 1988). 
 Power spectra obtained for cases 
with $C_{jk} \rightarrow \gamma C_{jk}$
show that $\omega_1$ increases with
$\gamma$ much more slowly than the higher
frequency modes which is consistent with
the behavior observed in Figures 6 and 12. 
 For the conditions of Figure 15, the magnitude
$\omega_1$ is 
considerably smaller than that predicted
for the limit of strong self-gravity between
the rings where $\omega_1 \rightarrow
\sum L_j\omega_{pj}/\sum L_j$ [equation(47)].
 This limiting frequency
is the same as the frequency $\omega_t$
of Sparke and Casertano (1988).  
 In any case, 
the frequency $\omega_1$ is
so small (the corresponding period is $>8$ Gyr)
that it is irrelevant to the initial
value problems considered below.

 In the following four subsections we consider
different possible origins of warps in an
otherwise flat galaxy.  Subsection 3.1
discusses warp excitation by a passing
compact satellite;  \S 3.2
treats warp excitation by a compact sinking
satellite;  \S 3.3 treats the case of 
a tilted halo potential.
Subsection 3.4 considers the tilt evolution
for the case where the initial plane of
the disk material is tilted.

\subsection{Warp Excitation by a Passing Satellite}

 Consider the excitation of a warp in a galaxy
due to the passage of a satellite of mass
$M_s$ much less than the galaxy's mass.  The
satellite's orbit ${\bf r}_s(t)$ can easily
be calculated exactly for $M_s\rightarrow 0$, 
neglecting the back reaction of the perturbed galaxy
on the satellite.
 However, we assume that even
at closest approach the satellite
is far from the center of the galaxy
and therefore 
calculate the orbit in the halo potential,
equation (6a),  with the ellipticity of
the halo neglected for simplicity ($q=1$).
At the closest approach at 
${\rm min}(|{\bf r}_s|)\equiv r_{so}$, the 
satellite's speed is $v_{so}$, and
the angle between the satellite and the 
$(x,~y)$ plane is $\theta_{so}$.
Also, at closest approach, the
satellite is taken to be
at $({\bf r}_{s})_y =0$ and $({\bf r}_{s})_z >0$.
We consider both prograde and retrograde
satellite passages.

 In the presence of a satellite of finite 
mass $M_s$, the description of the galaxy
disk in terms of tilted circular rings 
breaks down.  The centers of the
rings are shifted from the origin,
and the rings become non-circular.
 For this reason we consider the
response of the galaxy to a 
`symmetrized' perturbation of
a satellite obtained by replacing
the actual satellite by two satellites,
each of equal mass $M_s/2$, with one
satellite in the orbit ${\bf r}_s(t)$
described above and the other in the
orbit $-{\bf r}_s(t)$.  With this 
prescription we then calculate the 
torque of the satellites on each ring
as a function of time,
$$
{\bf T}^s_j(t)={1\over2}M_j\int_0^{2\pi}
{ d\phi \over 2\pi}~ {\bf r}_j \times
\left[ - ~{\partial~ \over \partial{\bf r}_j} 
\Phi_s({\bf r}_j,{\bf r}_s)\right]
$$
$$+~ (~{\bf r}_s \rightarrow - ~{\bf r}_s~)~,
\eqno(59)$$
where ${\bf r}_j(t) =[r_j{\rm cos}\phi,~ r_j{\rm sin}\phi,~
h_j(\phi,t)]$, with $h_j(\phi,t)$ given by equation (8a),
is the position vector to a point on the $j$th ring, and
$j=1,..,N$ runs over the different rings.  
Here, $\Phi_s = -GM_s/[({\bf r}_j -{\bf r}_s)^2 +a^2]^{1\over2}$
is the gravitational potential of the satellite and $a$ is
its characteristic radius.  To obtain the
warp response of the galaxy we evaluate equations (59)
numerically for $j=1,..,N$ and include the complex
torques ${\cal T}_j^s(t) =T_{jx}^s(t)-iT_{jy}^s(t)$
on the right hand sides of
equations (55).

Figure 16a shows the Briggs plot of the
warp response $\theta(\varphi)$ of the galaxy
resulting from the retrograde passage of
a satellite of mass $M_s = 2 \times 10^{10} M_\odot$.  
 Note that the dependence of $\theta(\varphi)$ at
the times shown is of the form of a
{\it leading} spiral wave and that it is
qualitatively similar to the leading
spiral waves observed in many warped
galaxies (Briggs 1990). 
 Note also that the form of the spiral
wave is preserved for a long time. 
 However, the
amplitude of the warp $\theta$ is 
significantly smaller
than the warps
observed in many cases, and this agrees
with the conclusions of HT.  Nevertheless,
it is of interest to understand the 
behavior.  
 Figure 16b shows the warp
response for the retrograde case with and
without gravitational interactions ($C_{jk}$)
between the rings. 
  Clearly these interactions
give a strong phase-locking of the inner rings
of the disk (roughly, $r_1$ to $r_{14}$
or $10$ to $18$ kpc) which have the same ${\varphi_j}'s$ as
a function of time.  
 The phase-locking of two and three rings
was disscussed in \S 2.2 and \S 2.4.

\placefigure{fig16}

Figure 17a shows the Briggs plot for the
warp resulting from the prograde passage
of the satellite.  
 For early times ($t \leq 2$ Gyr) the form 
of $\theta(\varphi)$
is a {\it trailing} spiral wave which is
distinctly different from that for the
retrograde passage at a similar time.
 The amplitude
of the warp is noticably larger than for
the retrograde case but still smaller
than many observed warps.
 For later times ($t\ge 4$ Gyr) the
trailing wave `unwraps,' evolving
into a leading spiral wave of form
similar to that for the retrograde passage
for radial distances out to about $35$ kpc.
 For larger $r$, the $\theta(\varphi)$
curve has a `spur' where
 $\theta$ increase with $r$
while $\varphi$ remains roughly constant
(see \S 3.2).  This unwrapping appears
to be due to the phase-mixing of the 
fast precession modes which propagate
over the radial extent of the disk
in times less than $4$ Gyr.

\placefigure{fig17}

 The different behavior of the
prograde and retrograde cases for
early times ($t\leq 1$ Gyr) can be 
understood by considering the 
response of a single ring with
or without gravitational interactions
between rings.
 Figure 17b shows the
orbits $[\theta_{jx}(t),~\theta_{jy}(t)]$
of ring No. 25 (radius $r_{25}=25$ kpc)
including the gravitational interactions
for prograde and retrograde encounters.
 Consider the torque on the $j^{th}$ ring
${\cal T}_j^s=T_{jx}^s- i T_{jy}^s$ which 
enters on the right hand side of equation
(55).  The tilting of the ring during the
encounter is relatively small and can be
neglected for this discussion. 
 Consequently, $T_{jy}^s(t)$ is
an even function of $t$ if the 
pericenter is taken as $t=0$.  
 Further, $T_{jy}^s(t)$ is the
same for
prograde and retrograde encounters.
 For the considered geometry of the 
satellite orbit, $T_{jy}^s(0) < 0$,
and the magnitude of $T_{jy}^s(t)$
decreases monotonically with
increasing $|t|$.  
 On the other
hand, the $x-$torque $T_{jx}(t)$,
although smaller in magnitude, is
an odd function of time, and
it has opposite signs for prograde
and retrograde encounters.
 Very roughly 
$T_{jx}^s(t) \propto d[T_{jy}^s(t)]/dt$ 
for the prograde
case whereas
$T_{jx}^s(t) \propto - d[T_{jy}^s(t)]/dt$
for the retrograde case.
 As a function of time, ${\cal T}_j^s$
traces out a closed {\it clockwise} 
path in the complex
plane for the prograde case, whereas
this path is {\it counterclockwise} for the
retrograde case.
  For the prograde case, the path of
${\cal T}_j^s$ is in the same clockwise
sense as the path of the orbit ${\Theta_j}=
\theta_{xj}-i\theta_{yj}$ for the {\it fast}
precession mode (see \S2.1), whereas
for the retrograde case the path of 
${\cal T}_j^s$ is in the same direction of
motion as $\Theta_j$ for the {\it slow}
precession mode.  For this reason a prograde
encounter couples strongly to the fast
precession mode, whereas a retrograde
encounter couples mainly to the slow
precession mode.

\subsection{Warp Excitation by a Sinking Satellite}

Here, we discuss the behavior of warps excited
by a slowly sinking compact minor satellite. 
 The satellite is assumed to be minor
in the respect that its mass 
is much smaller than that of
the galaxy plus halo matter within
say $30$ kpc. 
 The sinking of a more massive
satellite is likely to cause
substantial thickening of the
disk (Walker, Mihos, and Hernquist 1996)
which is not observed.
 Initially, the
satellite is assumed to be in an approximately
circular bound orbit in the halo gravitational
potential  
at a large radius $r_{so}$ in
the $(x,z)$ plane at an ange $\theta_{so}$ above
the $(x,y)$ plane.  
 The satellite slowly sinks
owing to dynamical friction with the halo matter
[equation (6a) with $q=1$]
which in the simplest description (BT, p. 428)
causes the specific angular momentum of the
satellite $\ell_s \approx r_s(t) v_o$ to decrease as
$$ {d\ell_s \over dt} \approx -0.428{G M_s 
\over r_s(t)} ~\ell n \Lambda~,
\eqno(60a)$$
where $\ell n \Lambda$ is the Coulomb logarithm.
 The time for the satellite to sink from $r_{so}$
to the galaxy's center is
$$t_{sink} \approx4.40~{\rm Gyr}\left({10^{10} M_\odot
\over M_s}\right)
\left({r_{so}\over 50~{\rm kpc}}\right)^2 \times
$$
$$
\left({v_o \over 200~{\rm km/s}}\right)
\left({3 \over \ell n \Lambda}\right)~.
\eqno(60b)$$
The disk response is found by
integrating equations (55) including
the numerically evaluated
torques (59) for ${\bf r}_s(t)$ at 
each time step for a `symmetrized'
sinking analogous to the approach
discussed in \S 3.1.

 Figure 18 shows the Briggs plots
of the warp response $\theta(\varphi)$
of the galaxy at different times
after the sinking (${\bf r}_s =0$) of a
satellite of mass $M_s = 10^{10} M_\odot$ 
in retrograde and prograde orbits.
A number of points are observed:

\placefigure{fig18} 

(1) The
warp amplitudes are larger than for the 
passing satellite 
(of mass $M_s = 2 \times 10^{10} M_\odot$)
discussed in \S 3.1, but the amplitudes are still
smaller than some observed warps (Briggs 1990).

(2) Coiling in the $\theta(\varphi)$
curves at early times ($t\leq 2$ Gyr) 
tends to disappear at later times ($t \geq4$ Gyr).
 We believe but have not proven that this is
due to the phase-mixing of the fast precession
modes which propagate over the radial extent of
the disk in times $< 4$ Gyr (see also \S 3.1).

(3) For both the retrograde and 
prograde cases
$\theta(\varphi)$ exhibits 
a leading spiral wave
for $t \geq 4$ Gyr qualitatively 
of the form
observed for M 83 (see Figure 2c).
However, the tilt amplitude 
$\theta$ is significantly
smaller than that for M 83.

The dissipative torques due to dynamical
friction [equation (56)] and that due to relative
friction [equation (57)] have different
consequences for the warp evolution. 
 For a dynamical friction coefficient
$\beta_j/\Omega_1=0.1$, the warp
amplitudes $\theta$ for the cases
of Figures 18a and 18b are reduced by
factors $\sim 0.5$ while the 
line-of-nodes angles are roughly the same.
 On the other hand, for relative friction
coefficients $\beta_j^\prime/\Omega_1 \leq 1$,
short radial wavelengths are damped out
without appreciably affecting the overall shape of
the $\theta(\varphi)$ curves.

\subsection{Tilted Halo Potential}

 Dekel and Shlosman (1983) and Toomre (1983)
proposed that observed warps of galaxies may
result from the fact that the dark matter
halo is oblate and is 
rigidly tilted with respect
to the inner disk.  
 To consider this possibility 
we generalize equation (55) to the case of
$N$ tilted rings in a halo 
potential which is
rigidly tilted by an angle, 
say, $\Theta_h = \theta_{hx}$
(with $\theta_{hx}^2 \ll1$) with respect to
the $z-$axis of the inner disk.  We find
$$I_j\left({d^2 \over dt^2}+2i\Omega_j
{d\over dt}\right)\Theta_j=-\Delta_{dj}^2I_j\Theta_j$$
$$
-\Delta_{hj}^2I_j(\Theta_j-\Theta_h)-\sum_{k \neq j}^N
{ C_{jk}}\left(\Theta_j -\Theta_k\right),
\eqno(61)$$
where $\Delta_{dj}^2 
\equiv \kappa_{idzj}^2-\Omega_{idj}^2$
and $\Delta_{hj}^2 
\equiv \kappa_{hzj}^2-\Omega_{hj}^2$.
The damping terms are still 
given by equations (56)
and (57).

 For a special disk tilt, $\Theta_j^L 
={\rm funct.}(j)$, the right-hand-side
of equation (61) is zero so that $\Theta_j^L$
is time-independent.  This $\Theta_j^L$ 
corresponds to the Laplacian surface of the
disk (BT, p. 413).  It is given by 
$$\sum_{k=1}^N {\cal M}_{jk}\Theta_k^L 
= \Delta_{hj}^2\Theta_h~,
$$
$${\cal M}_{jk}=\bigg(\Delta_{dj}^2+\Delta_{hj}^2+
{1\over I_j}{\sum_\ell}^\prime C_{j\ell}\bigg)\delta_{jk}
-{1\over I_j}C_{jk}(1-\delta_{jk})~,
\eqno(62)$$
where $\delta_{jk}$ is the Kronecker delta
and the prime on the summation means that
the diagonal terms are omitted.  
 The damping terms (56) and (57) are of course zero
for the Laplacian tilt.  
 If the self-gravity of the rings ($C_{jk}$) is
negligible, $\Theta_j^L=[\Delta_{hj}^2/
(\Delta_{hj}^2+\Delta_{dj}^2)]\Theta_h$.
 Equation (62) can always be inverted
to give $\Theta_j^L$ if $\Delta_j^2 >0$.
This is because the $N-$ring
system is stable for $\Delta_j^2>0$ (see discussion
following equation (53b)) which
implies ${\rm det}({\cal M}_{jk}) >0$.  

 A general solution of equation (61) can
evidently be written as $\Theta_j(t) = \Theta_j^L
+ \Theta_j^\prime(t)$, where $\Theta_j^\prime$
obeys equation (55) which has no reference
to the halo tilt.  Over a sufficiently long
period of time, damping given by equation
(56) and/or (57) will give $\Theta_j^\prime(t)
\rightarrow 0$ as discussed in 
the paragraph after next.  

\placefigure{fig19}

 Figure 19a shows the Laplacian tilt
$\Theta_j^L$ for a representative case with
and without the self-gravity 
of the rings.
 Figure 19b shows the Briggs
plots $\theta(\varphi)$ 
at times $t=4$ and $8$ Gyr for
a disk started from a Laplacian tilt
and from a $20\%$ deviation from a 
Laplacian tilt.  
 The Briggs plot for the Laplacian
tilt is in contrast with
the leading spiral wave observed,
for example, in M 83 (Figures 1 and 2)
(Briggs 1990).  However, the 
Laplacian tilt may be relevant
to cases such as NGC 3718 which
shows a relatively straight line-of-nodes.
  
 There is no 
evident reason for a disk to
be initially `set up'
in a Laplacian tilt.  
Figure 19b shows that deviations
from the Laplacian tilt evolve
to give a line-of-nodes which is
not straight.
 As mentioned, dissipative torques due to
dynamical friction [equation (56)]
and/or relative friction [equation(57)]
act to damp out the deviation 
($\Theta_j^\prime$) from the Laplacian
tilt over a period of time.
 For example, for $\beta_j/\Omega_1 =0.1$
in equation (56) and no relative friction,
the maximum line-of-nodes 
deviation ($\varphi_{49}$) at $8$ Gyr is 
reduced by a factor $\sim 0.4$ from
the case with no dissipation 
shown in Figure 19b.  
 On the other hand for $\beta_j^\prime/\Omega_1
=1.0$ and no dynamical friction, the maximum
line-of-nodes deviation is reduced by a
factor $\sim 0.5$. 

\subsection{Initially Tilted Disk Plane}

Here we consider the possibility that at some
intial time an outer gaseous 
disk is formed which is tilted
with respect
to the plane of the inner disk.
 This situation could arise by the
capture, tidal-disruption, radial-spreading,
and cooling of a low mass gas cloud by a
disk galaxy.
 For a spherical cloud of mass $M_{cl}$ and
radius $a_{cl}$ in an approximately 
circular orbit of radius $r$, 
tidal breakup occurs roughly for
$a_{cl} \geq \sim {2\over 5}(GM_{cl} r^2/v_o^2)^{1\over 3}$
$\approx 3{\rm kpc}
(M_{cl}/10^{10}M_\odot)^{1\over3}
(r/20{\rm kpc})^{2\over 3}~~$ 
$(200{\rm km/s}/v_o)^{2\over 3}$,
where $v_o$ is the circular velocity
in the flat rotation region of the 
galaxy [equation (6b)].
 Treatment of the cloud disruption
is beyond the scope of the present
work.  
 Instead, we study the
initial value problem where at $t=0$
there is a smooth transition from 
an untilted inner disk to a tilted
outer disk with a straight 
line-of-nodes.  Specifically, we
take
$$
\Theta_j(t=0)={\theta_{xo}\over 2}\bigg\{1-{\rm cos}
[\pi(r_j-r_1)/(r_N-r_1)]\bigg\}~,
\eqno(63a)$$
$${d\Theta_j(t=0)\over dt} =0~,
\eqno(63b)$$
where $\theta_{xo}$
is the initial tilt angle of the
outer disk. 
 Similar behavior is found for other smooth
variations of $\Theta_j$.  The smooth
variation of $\Theta_j$ avoids the 
excitation of short radial wavelength
modes.  

\placefigure{fig20}

 Figure 20 shows the nature of the warp
evolution resulting from the initial
conditions (63). 
 Figure 21 shows a surface plot
of the disk at $t=2$ Gyr. 
 A leading spiral
wave appears in the 
Briggs plot (panel {\bf c}),
and its amplitude grows with time.
 Progressively, the interior region
of the disk (say, $10 \leq r \leq 25$ kpc)
flattens to the plane of the inner disk
while the exterior region warp amplitude
$\theta$ grows.
 The dependences shown in Figure 20
for $t=2-4$ Gyr are
{\it qualitatively} similar to
those observed for M 83
(see Figure 2).  The outer part of 
the disk 
shows an inverse power law dependence of the 
line-of-nodes $\varphi$ on $r$ 
(panel {\bf b} of Figure 20) which is
qualitatively similar to that for
M 83 where $\varphi \propto r^{-1.5}$. 

\placefigure{fig21}

The dissipative torques due to dynamical
friction [equation (56)] and that due to
relative friction [equation (57)] affect
the results of Figure 20 in different ways.
For a dynamical friction coefficient
$\beta_j/\Omega_1=0.1$, and no relative
friction, the maximum warp amplitude 
at $t=4$ Gyr is reduced by about $30\%$.
On the other hand, for no dynamical friction, but
a relative friction coefficient $\beta_j^\prime=
1.0$, the warp at $t=4$ Gyr is essentially unchanged.

\section{Continuum Limit}

 In the continuum limit, the Hunter and Toomre (1969)
equation for the vertical displacement of
the disk $h(r,\phi,t)$
can be obtained
from equation (55) by recalling that 
$h=
r\theta_x(t){\rm sin}\phi-r\theta_y(t){\rm cos}\phi$
$={\cal I}m[r ~\Theta ~{\rm exp}(i\phi)]$.  We
omit for the moment the damping terms (56) and (57).
Multiplying equation (55) by $r~{\rm exp}(i\phi)$
and taking the imaginary part gives
$$ I_j\left({\partial \over \partial t} + 
\Omega_j{\partial \over \partial \phi}\right)^2 h_j
= -~I_j\left(\Omega_j^2+\Delta_j^2\right)~ h_j ~~~~
$$
$$~~~~~-\sum_{k\neq j}^N C_{jk}~ h_j+\sum_{k\neq j}^N C_{jk}~ 
r_jh(r_k,\phi,t)/r_k~,
\eqno(64)$$
where the left hand side of this equation 
follows from the same steps as for equation (21),
where $h_j = h(r_j,\phi,t)$, and where
$I_j=2 \pi r_j^3 \Delta r \Sigma(r_j)$ with
$\Delta r$ taken as a constant for simplicity. 
 The continuum limit is $N \rightarrow \infty$ 
and $\Delta r \propto 1/N \rightarrow 0$.

 From the definitions (14) and (18), we
have on the right hand side
of equation (64),  $\Omega_j^2+\Delta_j^2 =
\bar \kappa_{jz}^2+\Omega_j^2-\bar\Omega^2_j$.
From the decomposition of the potential (\S 1),
it is clear that $\Omega_j^2-\bar\Omega_j^2$
is the contribution to the disk rotation 
from the outer disk; that is,
$$\Omega_j^2-\bar\Omega_j^2 =
{1\over r}{\partial \Phi_{od} \over \partial r}=
$$
$${G\over r} \int d^2 r^\prime~
{\Sigma(r') ~[r - r^\prime {\rm cos}(\phi') ] \over
[r^2+(r^\prime)^2-2rr^\prime {\rm cos}(\phi') ]^{3\over 2}~},
\eqno(65)$$
where $d^2 r'= r'dr'd\phi'$, 
and where the integration is over the outer part of the 
disk, $r\geq r_1$, with $r_1$ chosen 
to be significantly larger than
the scale-length $r_d$ of the inner
disk (see \S 2).  Note that the summations
in equation (64) are also over the outer
part of the disk.  Thus, one term which 
contributes to the right hand side of equation
(64) is
$$-I_j(\Omega^2-\bar\Omega_j^2)h_j=
-~GI_j h_j\int d^2r^\prime~{\Sigma^\prime~ [r - r'{\rm cos}(\phi')]
\over r~ [.~.~.]^{3\over2}}~,
\eqno(66a) $$
where $\Sigma' \equiv \Sigma(r')$, and
the factor in the denominator to the $3/2$ power is
the same as in equation (65).  Using the
definition of $C_{jk}$ in equation (34), the first 
sum on the right hand side of equation (64) can
be converted to an integral in the continuum limit,
$${\sum}_1= -~GI_j h_j \int d^2r'~{\Sigma^\prime~ r'~{\rm cos}
(\phi')
\over r~[.~.~.]^{3\over2}}~.~~~~
\eqno(66b)$$
Similarly, the continuum limit of the second
sum in equation (64) gives
$${\sum}_2 = 
GI_j\int d^2r'~{\Sigma'~h(r',\phi,t)~{\rm cos}
(\phi-\phi') 
\over [r^2+(r')^2-2rr'{\rm cos}(\phi-\phi')]^{3\over2}}~,
$$
$$
~~~~~~~~=GI_j\int d^2r'~{\Sigma'~h(r',\phi',t)
\over [r^2+(r')^2-2rr'{\rm cos}(\phi-\phi')]^{3\over2}}~,
\eqno(66c) $$
in view of equation (31c).
 
 Combining equations (64)-(66) and dropping the $j$
subscripts gives
$$ \Sigma \left({\partial \over \partial t}
+\Omega {\partial \over \partial \phi} \right)^2 h =
-~\Sigma ~ \bar \kappa_z^2~ h
$$
$$-~G\Sigma  \int d^2r' {\Sigma'~(h-h') 
\over [r^2+(r')^2-2rr'{\rm cos}(\phi-\phi')]^{3\over 2}}~,
\eqno(67)
$$
where $h'\equiv h(r',\phi',t)$.  Note that $\bar \kappa_z$
includes the vertical restoring force of the inner
disk and the halo.   Thus, equation (67) is the same
as Hunter and Toomre's (1969) equation for $h$, in that they
did not separate out the inner disk and did not include
a halo potential.

 The Newtonian drag term (56) due to dynamical 
friction gives a contribution
$$ - ~\beta ~\Sigma~  {\partial h \over \partial t}
\eqno(68)$$
on the right hand side of equation (67).
The damping term (57) due to relative friction 
between the rings gives a contribution
$${1\over r^2}{\partial \over \partial r} \left[ \nu_t~ \Sigma~
r^3 {\partial \over \partial r} \left( {1 \over r}
{\partial h \over \partial t} \right) \right ]
\eqno(69)$$
on the right hand side of equation (67).

Multiplying equation (67) by $2d^2r 
(\partial h/\partial t)$, integrating over the outer
disk, and following the steps of HT (their
Appendix B) gives, in the absence of 
dissipation, the constant of the motion
$${\cal E}_{warp}=
\int d^2r~\Sigma\bigg[
\bigg({\partial h \over \partial t}\bigg)^2+
\Delta^2 h^2 \bigg]
$$
$$+{G\over 2}\int_0^\infty dr~r^2~\Sigma
\int_0^\infty dr'~(r')^2~\Sigma'\int_0^{2\pi} d\phi~ \times
$$
$$
J(r,r')\left[{h(r',\phi,t)\over r'}-
{h(r,\phi,t)\over r}\right]^2~,
\eqno(70)$$ 
where $J(r,r')$ is given by equation (30b).   
 Substituting $h(r,\phi,t)=r\theta(t){\rm sin}[\phi-\varphi(t)]$
from equation (8a),
one readily finds that ${\cal E}_{warp}$ is identical to
${\cal E}_{rings}={\cal H}$ of equation (53b).  Note that
${\cal E}_{warp}$ is non-negative 
and therefore the disk is stable
if $\Delta^2(r) \geq 0$.

Including the dissipation terms (68) and (69) gives
$${d {\cal E}_{warp} \over dt}=
-2\int d^2r~\beta~ \Sigma 
\left({\partial h \over \partial t}\right)^2
$$
$$-~2\int d^2r~ \nu_t~ \Sigma 
\left[r{\partial \over \partial r}
\bigg({1 \over r} {\partial h \over 
\partial t} \bigg) \right]^2~.
\eqno(71)$$
The relative friction is of course 
zero if $h/r$ is a function
only of time which corresponds to a rigid tilt.

The second constant of the motion ${\cal P}_{rings}$,
found in \S 2.1 and \S 3 [equation (53c)], can of course
be expressed in terms of $h(r,\phi,t)$ in the continuum
limit.  Instead, we obtain this new constant directly
from the HT equation for $h$ by multiplying equation (67)
by $2d^2r (\partial h/\partial \phi)$ and integrating
over the outer disk.  This gives
$${\cal P}_{warp}= 2\int d^2r~\Sigma~ 
\bigg[\Omega\left({\partial h
\over \partial \phi} \right)^2 +
{\partial h \over \partial \phi}{\partial h 
\over \partial t}\bigg]
={\rm Const.}
$$
$$=\int d^2r~\Sigma ~r^2 \theta^2 (\Omega -
\dot \varphi)~,~~~~~~~~~~~~~~~
\eqno(72)$$
and ${\cal P}_{warp} ={\cal P}_{rings}$ 
in the continuum limit, 
in the absence of dissipation. 
 With dissipation included, we
have not found a simple 
expression for $d{\cal P}_{warp}/dt$.

\section {Conclusions}

This paper develops a 
representation for the
antisymmetric small amplitude warp
dynamics of a
self-gravitating disk 
in terms of $N$ tilted
concentric rings.  
 That is, we consider only
azimuthal mode number of 
the warp $m=1$ (or $-1$).
 This representation is suggested by the 
kinematic ring model of 
Rogstad et al. (1974)
which is used in interpreting 
HI disk warps (Briggs 1990). 
 The rings are considered to be
in a fixed oblate (or prolate)
halo potential and in the potential of an
inner, untilted disk. 
 Different initial value problems
are studied using our $N-$ring model.

We first consider in detail the 
tilting dynamics of one ring in the
potential of the halo and inner disk.
The equations of motion are shown
to have a particularly simple form when
written in terms of the complex
tilt angle $\Theta = \theta {\rm exp}(-i\varphi)$,
where $\theta$ is the actual tilt angle,
and $\varphi$ is the angle of the line-of-nodes
relative to the inertial $x-$axis.   
 The single ring has a slow and a fast precession
mode which are analogous to the normal
modes of vibration of a non-rotating
mechancial system.  For the slow (fast)
mode $\Theta(t)$ rotates in the clockwise
(counter clockwise) direction in the complex
plane.  The dynamical equation for $\Theta$
of course has an energy constant of the motion,
${\cal E}_{ring}$.
Additionally, we show that there is a second
constant of the motion.  
 By writing out the Lagrangian for one ring,
${\cal L}={\cal L}(\theta,\dot \theta, \dot \varphi)$,
this constant of the motion is recognized as the
canonical angular momentum for the $\varphi$
coordinate, $P_\varphi = \partial {\cal L}/\partial 
\dot\varphi =$ Const.

 We examine the influence of the drag on
the motion of one ring
through the halo matter
due to dynamical friction. 
 The
full calculation of this drag (Nelson and
Tremaine 1995) is beyond the scope of this work.
Instead, we include a linear drag torque
in the equation of motion for $\Theta$ 
which we estimate following the approach
of Weinberg (1985).  Inclusion of this term
shows that dissipation can destabilize the 
ring tilting in a prolate halo potential.
 The instability is of the negative energy
type;  that is, it occurs only for ${\cal E}_{ring} < 0$.

 We next consider the particularly interesting
case of two tilted, gravitationally interacting
rings of different radii
in the potential of the halo
and inner disk.  We derive dynamical equations
for $\Theta_1 = \theta_1 {\rm exp}(-i\varphi_1)$
and $\Theta_2 = \theta_2 {\rm exp}(-i\varphi_2)$.
This case shows the phenomenon of phase-locking
of the line-of-nodes of the two rings.
The gravitational interaction between the
two rings is appropriately measured by the
coupling strength $\xi_{12}$ of equation (48).
 As $\xi_{12}$ increases the ring motion 
changes from that of independent precession
for $\xi_{12} < 1$ to phase-locked precession
for $\xi_{12}>1$ where the lines-of-nodes
coincide approximately,
$\varphi_1 \approx \varphi_2$.
 This phase-locking persists for the case
of $N>2$ rings.

 We examine the influence of dissipative
forces on the tilting motion of two rings.
 In addition to dynamical friction, which is
included as linear drag torques on each ring,
we include the `relative' friction which 
can result from the differential vertical motions
of two gaseous rings due to turbulent viscosity.
 The corresponding drag torques on each
ring have a definite form owing to
the requirement that the total
angular momentum of the two rings
be conserved in the absence of external
torques. 
 The turbulent viscosity is assumed to be 
described by the $\alpha$ model 
of Shakura (1973) and Shakura 
and Sunyaev (1973). 
 We find that the
relative friction, as well as the dynamical
friction, can destabilize the motion of two
rings in a prolate halo potential.

 We study the case of two tilted
rings which rotate in opposite
directions.  This situation, which is
pertinent to observed counter-rotating
galaxies (see, for example, Jore, Broeils, and
Haynes 1996),
is found to be unstable in a prolate
halo potential for certain conditions 
in the absence of dissipation.  
 That is,
a dynamical instability may occur.

 For the general case of $N-$tilted,
gravitationally interacting rings
of radii $r_j=r_1+(j-1)\Delta r$ in the
potential of the halo and inner disk we
obtain equatins of motion for $\Theta_j=
\theta_j{\rm exp}(-i\varphi_j)$ for $j=1,..,N$.
The Lagrangian has the form ${\cal L}=
{\cal L}(\theta_j, \varphi_j, \dot \theta_j,
\dot \varphi_j)$, and this gives, in the 
absence of dissipative torques, both an
energy constant of the motion and a total
canonical angular momentum constant of the motion,
$P_\varphi = {\sum }_j \partial {\cal L}/\partial
\dot \varphi_j =$ Const.  We comment on
the numerical solutions for $\Theta_j$ for
different values of $N$, and we give a
sample power spectrum for $N=49$ which shows
the distribution of the $2N$ mode frequencies.
We argue that for treatment of intial
value problems of time duration $t_{max}$, 
mode frequency differences $\delta \omega$
smaller than $1/t_{max}$ are irrelevant.  This
leads to the estimate $N = {\cal O}(10^2)$ for
the number of rings needed to treat warps
in galactic disks.

In the continuum limit $N \rightarrow \infty$,
the ring model is shown to give the Hunter
and Toomre (1969) dynamical equation for
the vertical displacement of the disk $h(r,\phi,t)$.
Dissipative torques due to dynamical friction
and/or relative friction (for gaseous rings)
are shown to
cause the energy of the warp ${\cal E}_{warp}$
to decrease with time.

We have numerically solved the 
dynamical equations for
$\Theta_j$ for $N=49$ for four
different types of initial conditions
which may give rise to observed warps in
galaxies:  
 (1) warp excitation by a passing
satellite with relatively large impact
parameter; 
 (2) excitation by a 
slowly sinking compact minor satellite;
 (3) warp evolution in a tilted halo
potential (Dekel and Shlosman 1983;  and
Toomre 1983);
 and
(4) warp evolution resulting from an 
initially tilted disk plane due to the
tidal breakup of a gas cloud.
 The nature of the disk response
at different times is most clearly
shown in the polar plots of $\theta_j$
versus $\varphi_j$ which we refer to
as Briggs plots.  
 This is one of the
forms used by Briggs (1990) to 
describe the warp geometry as
deduced by fitting a kinematic
ring model (Rogstad et al. 1974) to
HI observations.

 For case (1) we find that the polar
plots $\theta_j(\varphi_j)$ have the
shape of a leading spiral wave qualitatively
of the form observed, but the warp 
amplitude $\theta_j$ is smaller than
observed in many galaxies.  
 The small warp amplitudes found here are
in agreement with the conclusions of
HT.  
 The leading spiral in the $\theta_j(\varphi_j)$
curves in the outer part of the disk
(where the self-gravity between the rings is small) 
results from the dominance of the slow
precession mode which gives very roughly
$\dot \varphi_j \sim \omega_{pj} \propto
-1/r$, where $\omega_{pj}$ is the single
ring slow precession frequency.
 The fast precession modes disappear in a
few Gyrs due to phase-mixing.

 For case (2) of a sinking satellite, we
find that the $\theta_j(\varphi_j)$
curves have a large noise component and are unlike
the observed curves in the absence of relative
friction between the rings.
 This is due to the violent excitation from
the satellite passing through the disk.  
 However, with
relative friction included, the $\theta_j(\varphi_j)$
curves are smooth leading spiral waves qualitatively
similar in shape to the observed curves.  
 The warp
amplitude, although significantly larger than for
case (1), is still smaller than that for many
observed warps.
  We find that the inner part of the disk where
$\theta_j$ is small has an approximately straight 
line-of-nodes ($\varphi_j$ independent of $j$)
for both cases (1) and (2) as observed by
Briggs (1990) (his `rule of behavior' No. 2).
 We show that this is due to the 
above-mentioned phase-locking of the 
lines-of-nodes of the inner rings of the
disk due to self gravity between the rings.

 For case (3) of an oblate halo potential
rigidly tilted with respect to the inner
disk of the galaxy, there is a unique
time independent Laplacian warp
$\Theta_j = \Theta_j^L$ (BT, p. 413)
which has a straight line-of-nodes 
($\varphi_j=$ Const.). 
 However, it is unlikely that a galactic
HI disk is `set up' with the Laplacian
tilt.  Deviations from $\Theta_j^L$
evolve to give a line-of-nodes which is
not straight.  Over the age
of the galaxy, the deviations
from $\Theta_j^L$ may damp out due to
dynamical friction and/or relative
friction between the rings.  This
case is possibly relevant to the 
warp of NGC 3718 which shows an
approximately straight line-of-nodes
out to $r \approx 1.75 R_{Hol}$ (Briggs 1990).

 Case (4) of an initially tilted disk plane
with a straight line-of-nodes gives: 
(1) polar plots $\theta_j(\varphi_j)$ for 
$t\sim 2 - 6$ Gyr with the leading spiral shape
qualitatively of the form observed;  
 (2) straight line-of-nodes ($\varphi_j \approx$
Const.) in the inner part of the disk as observed
as a result of phase-locking 
due to self-gravity; and
(3) a warp amplitude linearly 
dependent on the initial tilt of
the disk plane.  
 Further, the line-of-nodes angle 
$\varphi(r)$ in the outer part 
of the disk is found to have an inverse
power law dependence on $r$ qualitatively
of the form of shown for M 83 (see Figure
2b).

\newpage

\acknowledgments{We thank M.P. Haynes for pointing
out the paper of F.H. Briggs (1990) and for valuable
discussions which stimulated this work.
Further, we thank F.H. Briggs, G. Contopoulos, 
and M.M. Romanova for valuable discussions.  
We thank
I.L. Tregillis for critical checking 
of an early version of this work. 
This work was supported in part 
by NSF grant AST-9320068.}

\newpage

\figcaption{Surface plot of the warped disk
of M 83 (NGC 5236) obtained 
from the data of Briggs (1990).
The rotation of the galaxy is 
clockwise so that the
warp of the disk is a 
{\it leading} spiral wave (Briggs 1990).
The scales are in units of 
$10$ kpc for a Holmberg radius
of M 83 of $R_{Hol}=7.3$ 
arcmin and a distance of M 83 of
$5.9$ Mpc which assumes 
a Hubble constant $H_o=65$ km/s/Mpc
(Briggs 1996, private communication).
The colors are in $16$ steps uniformly
spaced between $z_{min}$ and $z_{max}$. 
\label{fig1}}

\figcaption{Nature of the 
warp of the disk of M 83
from Briggs (1990).
The top panel ({\bf a}) shows 
the radial ($r$) dependence of the 
tilt angle $\theta(r)$. 
The middle panel ({\bf b}) shows
the $r$-dependence of the angle 
of the line-of-nodes $\varphi(r)$.
The bottom panel ({\bf c}) shows the
dependence of $\theta$ on $\varphi$, which we refer
to as a Briggs plot.  
 The radii are in units of $10$ kpc for
the parameters mentioned 
in the caption of Figure 1.
 In these units the Holmberg radius
is $R_{Hol}\approx 1.26$.
 In panel $({\bf b})$, the dashed
curve shows the least-square-fit of a power law
dependence to the curve for $1 \le r \le 3.4$ which
gives $\varphi \approx 216^o r^{-1.48}$ 
(with an $R$ value of $0.993$).  
Note that in the Briggs plot
$\varphi$ has been taken to increase 
in the clockwise direction
to account for the fact 
that the galaxy rotates in
the clockwise direction.  
 Except for our
dicussion of counter rotating rings, we assume
that the matter of the rings (representing
the galaxy disk) rotates in the counter-clockwise
direction. 
Note that the data points given here
have only an approximate correspondence
with those given by Briggs in that
we have scanned and digitized 
the data from his figures.
\label{fig2}} 

\figcaption{Geometry of one tilted 
 ring of a disk galaxy. 
For the case shown, the line-of-nodes 
is along the $x-$axis, that is,
$\varphi=0$.  Also, $\theta_x > 0$,
$\theta_y=0$, and ${\bf n}$
denotes the normal to the ring
plane. \label{fig3}}

\figcaption{Illustrative single ring
orbits for a case shown in ({\bf a}) where the slow mode
is dominant ($C_1=0.2,~C_2=0.8$) and in ({\bf b}) where
the fast mode is dominant ($C_1=0.8,~C_2=0.2$).  For
both plots, $i$ corresponds to the starting point,
and $f$ to the end point at $t_f = 62.8/\Omega$,
and $\Delta^2/\Omega^2 =0.2$.
\label{fig4}}

\figcaption{Dependence of the dimensionless integral
$J_{12}^\prime$ defined in
equation (45e) on $\delta = r_1/r_2$. 
The dashed lines correspond to
approximations discussed in the
text. \label{fig5}}
 
\figcaption{The top panel ({\bf a}) shows the
dependences of the two slow 
precession frequencies of
two co-rotating rings in an {\it oblate} halo on the 
strength of the gravitational interaction
measured by $\omega_{g12}$ 
obtained the full equation (43)
and the approximate equation (44).  The
frequencies are measured in units of $\Omega_1$.   
 For this figure,  
$\Omega_1/\Omega_2=2$, 
$I_1/I_2=1$, and 
$\Delta_j^2=0.2\Omega_j^2$.   
 [Note that the two fast
precession modes, $\omega_3$ and $\omega_4$, 
have frequencies $1.05~ {\rm and}~ 2.1$ for $\omega_{g12}=0$,
and $1.19 ~{\rm and}~ 2.19$ for $\omega_{g12}=0.1$.
The $\omega_3$ mode has $\Theta_2/\Theta_1\gg 1$ and thus
involves mainly motion of ring 2, whereas the $\omega_4$
mode has $-\Theta_2/\Theta_1 \ll 1$ and involves motion
mainly of ring 1.] \newline
$-$ The bottom panel ({\bf b}) shows 
the dependences of the amplitude 
ratios $\Theta_2/\Theta_1$
for the two slow precession
modes.
For small values of $\omega_{g12}$,
the $\omega_1$ mode involves 
motions mainly of ring 2,
whereas the $\omega_2$ mode 
involves motion mainly of ring 1. \label{fig6}}

\figcaption{Illustrative orbits 
$\{\Theta_1(t),~\Theta_2(t)\}$ for two
co-rotating rings in an oblate halo for
weak ({\bf a}, top panel) and strong 
({\bf b}, bottom panel) coupling for
the same parameters as Figure 6.
The initial conditions are $\Theta_1/\Theta_2=1/2$
and $\dot \Theta_1 =0=\dot \Theta_2$.
 The points labeled
by $f^\prime$s all correspond to the 
final time $t_f=30\pi/\Omega_1$ (which
is arbitrary).  The constants
of the motion ${\cal E}_{rings}$
and ${\cal P}_{rings}$ are constant
in the numerical integrations with 
fractional errors 
 $<~10^{-6}$.
 For the top panel, $\omega_{g12}/\Omega_1=
0.005$ ($\xi_{12}\approx 0.3$), 
and the rings precess almost
independently.  For the bottom panel,
$\omega_{g12}/\Omega_1 =0.015$ 
($\xi_{12}\approx 0.9$), and
the rings precess together with their 
lines-of-nodes phase-locked as shown in
Figure 8.  
 The dividing value separating
independent and phase-locked precession
is $\omega_{g12}/\Omega_1 \approx 0.010$ or
$\xi_{12} \approx 0.60$.  Of course, this
value depends on the inital value
$\Theta_1/\Theta_2$ for 
$\dot \Theta_j(0)=0$. 
\label{fig7}}

\figcaption{Azimuthal angle difference of
the line-of-nodes 
of two rings  $\varphi_2-\varphi_1$ for cases ({\bf a})
and ({\bf b}) of Figure 7.  At the final
time $t_f=30\pi/\Omega_1$, for the phase-locked
case $(\varphi_2-\varphi_1))_f \approx -24^o$,
whereas for the unlocked case $(\varphi_2-\varphi_1)_f
\approx 253.5^o$.
\label{fig8}}

\figcaption{The top panel ({\bf a}) shows the
dependences of the two slow 
precession frequencies of
two co-rotating rings in an {\it prolate} halo on the 
strength of the gravitational interaction
measured by $\omega_{g12}$ 
obtained equation (43).
The
frequencies are measured in units of $\Omega_1$.   
For this figure,  
$\Omega_1/\Omega_2=2$, 
$I_1/I_2=1$, and 
$\Delta_j^2=-0.2\Omega_j^2$.   
$-$ The bottom panel ({\bf b}) shows 
the dependences of the amplitude 
ratios $\Theta_2/\Theta_1$
for the two slow precession modes.
 For small values of $\omega_{g12}$,
the $\omega_1$ mode involves 
motions mainly of ring 2,
whereas the $\omega_2$ mode 
involves motion mainly of ring 1.
For large $\omega_{g12}$ the $\omega_2$
mode frequency approaches the limit 
given in equation (47) and $\Theta_2/\Theta_1
\rightarrow 1$.
\label{fig9}}

\figcaption{Region of instability of 
two rings, one co-rotating ($\Omega_1>0$)
and the other counter-rotating ($\Omega_2<0$), 
in a prolate halo
potential as discussed 
below equation (51). 
\label{fig10}}

\figcaption{The top panel ({\bf a}) shows the
dependences of the two slow 
precession frequencies of two
counter rotating rings ($\Omega_1>0$ and $\Omega_2<0$)
in an {\it oblate} halo on the 
strength of the gravitational interaction
measured by $\omega_{g12}$ 
obtained equation (43).
The
frequencies are measured in units of $\Omega_1$.   
For this figure,  
$\Omega_1/\Omega_2=-2$, 
$I_1/I_2=1$, and 
$\Delta_j^2=0.2\Omega_j^2$. \newline   
$-$ The bottom panel ({\bf b}) shows 
the dependences of the amplitude 
ratios $\Theta_2/\Theta_1$
for the two slow precession modes. \label{fig11}}

\figcaption{The top panel ({\bf a}) 
shows the dependence of the three 
slow precession frequencies of
a three ring system on the 
strength of the gravitational interaction
measured by $\omega_{g12}$. 
 The
frequencies are measured in units of $\Omega_1$.
For this figure, $(r_1,r_2,r_3)$ $=(1,2,3)$, 
$(\Omega_1,\Omega_2,\Omega_3)=
(1,{1\over2},{1\over3})$, 
$I_2/I_1=1,~I_3/I_2=1$, and 
$\Delta_j^2=0.2\Omega_j^2$.   
 For small values of $\omega_{g12}$,
the $\omega_1$ mode involves 
motion mainly of ring 3,
the $\omega_2$ mode 
motion of ring 2, and the $\omega_3$ mode
motion of ring 1.  For larger values of
$\omega_{g12}$ the motions of the different
rings become coupled.   For large $\omega_{g12}$
the slowest mode ($\omega_1$) approaches the
limiting value given by the generalization of
equation (47) to three rings.  
 [The three fast
precession modes have frequencies 
$0.698,~1.05,~ {\rm and}~ 2.09$ for $\omega_{g12}=0$
and $0.842,~1.30,~{\rm and}~ 2.20$ 
for $\omega_{g12}/\Omega_1=0.1$.]  
\newline
$-$ The bottom panel ({\bf b}) shows 
the $r$-dependence of the vertical
displacements $h$ (along
a line $\perp$ to the line-of-nodes) 
for the three
slow precession modes for
$\omega_{g12}/\Omega_1 = 0.1$.
 The vertical scale is arbitrary except
for the condition $h^2 \ll r^2$. 
 Smooth lines have been drawn through
the calculated values indicated by circles. 
\label{fig12}} 

\figcaption{Illustrative orbits 
$\Theta_j(t)$ for three
co-rotating rings in an {\it oblate} halo for
weak ({\bf a}, top panel) and strong 
({\bf b}, bottom panel) coupling for
the same parameters as Figure 12.
The initial conditions are $\Theta_j \propto r_j$
and $\dot \Theta_j=0$.
 The points labeled
by $f^\prime$s all correspond to the 
final time $t_f=45\pi/\Omega_1$ (which
is arbitrary).  
 For the top panel, $\omega_{g12}/\Omega_1=
0.005$ and the rings precess almost
independently.  For the bottom panel,
$\omega_{g12}/\Omega_1 =0.03$ and
the rings precess together with their 
lines-of-nodes phase-locked as shown in
figure 14.  
\label{fig13}}

\figcaption{Azimuthal angles $\varphi_j(t)$ of
the line-of-nodes for three rings 
for cases of:
({\bf a}) weak  
($\omega_{g12}=0.005\Omega_1$), ({\bf b})
intermediate  
($\omega_{g12}=0.015\Omega_1$),
and ({\bf c}) strong coupling ($\omega_{g12}=
0.03\Omega_1$). 
\label{fig14}}

\figcaption{The figure
shows the power spectrum as
a function of  
angular frequency $\omega$ 
of
$\theta_{x25}(t)$ - the 
$x-$component tilt
angle of the middle
ring at $r=25$ kpc - for a $N=49$
ring disk given a random initial
perturbation.  
 The
frequency $\omega$ is 
normalized by the angular velocity of the
disk at $r_1$, $\Omega_1$. 
 The spectrum is
obtained using a $1024$ point FFT.
  The initial perturbation
is taken to have a `white noise'
radial wavenumber spectrum;
that is, all radial wavenumber
modes are excited. 
The $\Theta_j(t)$ 
are obtained using a code which
solves equations (55) 
with the rings equally
spaced between $r_1=10$ kpc and $r_{49}=40$ kpc.
 Dissipation is neglected ($\beta_j=0=\beta_j'$).
  The mass of the inner disk in equation (4) is
$M_d=6 \times 10^{10} M_\odot$,
and the disk radial scale is $r_d=4$ kpc.
The halo potential is given by equation (6a)
with ellipticity $\varepsilon =0.0871$,
core radius
$r_o=15$ kpc, and circular velocity
$v_o = 200$ km/s. 
 The neutral hydrogen has a total
mass of $1.36 \times 10^{10}M_\odot$ and is
distributed according to equation (5) with
$R_H = 20$ kpc.
 The separation between low and high
frequency modes is roughly 
at $\omega/\Omega_1=0.75$  
 The lowest frequency of the spectrum is
$|\omega_1|/\Omega_1 \approx 0.0368$ while
the next two higher frequencies are
at  $\approx 0.0521$ and $0.0639$.
 Note that the limiting value of $\omega_1$ from
the generalization of equation (47) is
$\omega_1
\rightarrow \sum L_j\omega_{pj}/\sum L_j \approx
-~0.120 \Omega_1$ which is significantly 
larger in magnitude than the $\omega_1$ observed. 
\label{fig15}}

\figcaption{The top panel (${\bf a}$)
shows a polar plot (a Briggs plot) of the tilt angle
of the disk 
$\theta$ in degrees as a function of the angle
of the line-of-nodes $\varphi$ for the
case of a retrograde passage of a compact
satellite for times $t=1,~2,$ and
$3$ Gyr after the time of closest approach.
The plot shows a {\it leading} spiral wave
qualitatively of the form observed, but 
with a relatively small amplitude compared
with many observed warps (Briggs 1990).
\newline
$-$ The bottom panel (${\bf b}$) shows a comparison
of the disk response with and without gravitational
interaction ($C_{jk}$) between the rings.  This
figure shows the locking of the phase $\varphi$
in the inner part of the disk. 
\newline
$-~$
The results shown in ($\bf a$) and ($\bf b$)
are obtained from a code which
solves equations (55) including the
torques of equation (59) with $N=49$ rings equally
spaced between $r_1=10$ kpc and $r_{49}=40$ kpc.
 Thus the separation between rings 
is $\Delta r = 0.625$ kpc.  
The torques
are evaluated numerically at each time step. 
 The Newtonian drag is neglected, $\beta_j=0$
in equation (56).  
 A small relative friction is included, 
$\beta_j'/\Omega_1 =0.02$ in equation (57).
 [For a disk half-thickness $\Delta z =250$ pc
and sound speed $c_s=10$ km/s, this $\beta_j'$
corresponds to a value of
Shakura's (1973) viscosity parameter $\alpha
\approx 0.0624$.  
 This friction acts to smooth out sharp
`corners' which exist in the $\beta'_j=0$ curve
$\theta(\varphi)$.  
  Owing to this $\beta_j'$,
${\cal E}_{rings}$ decreases by $16$\% between
$t=1$ and $3$ Gyr.]
 The mass of the satellite is $2\times 10^{10}M_\odot$,
its core radius is $a=1$ kpc, 
and at closest approach it is at a distance
$r_{so} = 30$ kpc where it has a speed of
$400$ km/s and is located at an angle $30^o$ above
the $(x,y)$ plane.  
 The mass of the inner disk in equation (4) is
$M_d=6 \times 10^{10} M_\odot$,
and the disk radial scale is $r_d=4$ kpc.
The halo potential is given by equation (6a)
with ellipticity $\varepsilon =0.0871$,
core radius
$r_o=15$ kpc, and circular velocity
$v_o = 200$ km/s so that 
$(\kappa_{hz}^2-\Omega_h^2)/\Omega_h^2
=0.2$ for $r^2 \gg r_o^2$.  
 The mass
of halo matter inside of $r= 30$ kpc
is $7.4 \times 10^{10} M_\odot$.
The neutral hydrogen has a total
mass of $1.36 \times 10^{10}M_\odot$ and is
distributed according to equation (5) with
$R_H = 20$ kpc.
\label{fig16}}

\figcaption{
The top panel (${\bf a}$)
shows a polar plot of the tilt angle
of the disk 
$\theta$ in degrees as a function of the angle
of the line-of-nodes $\varphi$ for the
case of a prograde passage of a compact
satellite at times $t=1$ and $t=4$
Gyr after the time of closest approach.
 The other conditions are the same as given
in the caption to Figure 16.  The plot shows
that initially ($t = 1$ Gyr)
a {\it trailing} spiral wave
is formed.  
 Later, ($t=4$ Gyr) the trailing wave
`unwraps' and gives rise to a {\it leading}
spiral wave.
Analysis of
the evolution of the tilt excited by 
the prograde satellite shows that it is initially a
{\it fast} outward propagating wave with radial
phase velocity $v_{r}^{ph} \approx 24$ km/s
and radial wavelength $\lambda_r\approx 11$ kpc
for $r\sim 25$ kpc.  Roughly, $\lambda_r \propto r^{2.5}$.
\newline
$-$ The bottom panel (${\bf b}$) shows a comparison
of the orbits $[\theta_{25x}(t),~\theta_{25y}(t)]$ (in degrees)
of ring No. 25 at radius $r_{25}=25$ kpc for
$t=0-1$ Gyr for the cases
of retrograde and prograde satellite encounters
for the conditions of Figures (15a) and (16a).  
 The circles and squares are equally spaced at $0.1$ 
Gyr intervals.
 The ($i$) indicates $t=0$ and the ($f^\prime$s) correspond
to $t=1$ Gyr.  
 For the retrograde case, mainly the
slow precession mode of the ring is excited, whereas
for the prograde case mainly the fast precession
mode is initially excited.
\label{fig17}}

\figcaption{
The top panel (${\bf a}$)
shows a polar plot of the tilt angle
of the disk 
$\theta$ in degrees as a function of the angle
of the line-of-nodes $\varphi$ for the
case of a retrograde sinking of a compact
minor satellite of mass 
$M_s=10^{10} M_\odot$ at times $t=2$ and $t=4$
Gyr after the sinking (${\bf r}_s=0$).
 Initially, the rings are unperturbed
and the satellite is `turned on' at a distance
$r_s = 50$ kpc in the $(x,z)$ plane $30^o$ above
the $(x,y)$ plane.  The in-spiral of the satellite
is described by equation (60a) with
$\ell n \Lambda =3.$  The other conditions
are the same as described in the caption of 
Figure 16 except that the relative friction
is larger, $\beta_j^\prime/\Omega_1 = 0.2$.
The Newtonian drag is neglected.  
 This value of $\beta_j^\prime$ has the
effect of damping out short radial
wavelength features of the tilt while
not appreciably changing 
the overall $\theta(\varphi)$ curve.  
\newline
$-$ The bottom panel (${\bf b}$) shows the
Briggs plot for a prograde sinking for
conditions otherwise the same as in (${\bf a}$)
above.  
\label{fig18}}

\figcaption{The top panel ({\bf a}) shows
the Laplacian tilt angle of the rings
$\Theta_j^L = \theta_{jx}^L$ with and without
self-gravity of the rings ($C_{jk}$) for the
case where the halo potential [equation (6a)]
is rigidly tilted by an angle 
$\Theta_h=\theta_{hx} =
30^o$ with respect to the $z-$axis of the inner
disk of the galaxy. 
 The disk and halo are the same as in
Figure 16.
 The $\Theta_j^L$ are 
obtained by solving equation (62).
\newline
$-$ The bottom panel ({\bf b}) shows the
Briggs plots for the Laplacian tilt 
and that for a small initial deviation
from a Laplacian tilt at times $t=4$ and
$8$ Gyr.
 The initial deviation is taken to
be $\Theta_j= \theta_{jx}^L
\{1+0.2{\rm sin}[\pi(r_j-r_1)/2(r_N-r_1)]\}$.
 This deviation vanishes at $r_j=r_1$
and it has a maximum 
at the outer edge of the disk.
 Dissipative torques are neglected.
 This deviation from $\Theta^L_j$
gives rise to a leading spiral feature
at the outer edge of the disk at 
$t=8$ Gyr.  For a deviation with the
$+0.2$ factor replaced by $-0.2$ the
line-of-nodes angle at the outer
edge of the disk ($\varphi_{49}$)
at $t= 8$ Gyr 
has a similar magnitude to that 
in Figure 19b, but it is
negative.
\label{fig19}}
 
\figcaption{The top panel ({\bf a})
shows the radial dependence
of the tilt angle $\theta(r)$
for different times after the 
initial time ($t=0$) when
the disk tilt is give by
equations (63) with outer disk
tilt angle $\theta_{xo}=30^o$.
 The conditions of the disk
and halo are otherwise the
same as for Figure 16.
\newline
$-$The middle panel ({\bf b}) 
shows the radial dependence of
the line-of-nodes $\varphi(r)$.
 The constant value of $\varphi(r)$
in the inner part of the disk
results from phase-locking due
to the self-gravity of the rings
as discussed in \S 2.3 and \S 2.5.
 The dashed curves correspond to
least-square fits of a power law
to $\varphi(r)$. 
 For $t=2$ Gyr, the fit gives
$\varphi = 498^o r^{-1.42}$
for $r=23-40$ kpc (with an $R$
value of $0.96$).
 For $t=4$ Gyr, the fit gives
$\varphi=1680^o r^{-1.79}$
for $r=28-40$ kpc (with an
$R$ value of $0.97$).
 Note that for M 83 (Figure 2b),
the least-square fit gives
$\varphi=216^o r^{-1.48}$.
\newline
$-$ The bottom panel ({\bf c}) shows
the Briggs plots for a sequence of times.
A leading spiral wave forms, and
its amplitude grows with time.
The 
$\theta(\varphi)$ spiral for $t=2-4$ Gyr
 is qualitatively
similar to that
observed for M 83 (see Figure 2c).
 From results not shown here the similarity
continues up to $t\sim 6$ Gyr 
after which time $\theta$ and
$d\theta/dr$ become 
large at the outer edge of
the disk.  
 The present theory
assumes small tilting amplitudes 
$\theta^2 \ll 1$ for the 
linearization of the equations of motion.
The linearity 
allows an arbitrary rescaling $\Theta_j \rightarrow
K \Theta_j$ with $K=$ Const.  (or $\theta_j
\rightarrow K\theta_j$ and $\varphi_j
\rightarrow \varphi_j$) in Figure 20. 
\label{fig20}}

\figcaption{Surface plot of the warped
disk of Figure 20 at $t=2$ Gyr.   To facilitate
comparison with Figure 1 for M 83, the rotation of the
disk has been taken to be clockwise in this 
figure.  The warp is a leading
spiral wave the same as for M 83.
The colors are in $16$ steps uniformly
spaced between $z_{min}$ and $z_{max}$. 
\label{fig21}}

\end{document}